\documentclass[aps,twocolumn,superscriptaddress,longbibliography,floatfix]{revtex4-2}

\usepackage{amsmath,amssymb,bm}
\usepackage{graphicx}
\usepackage{booktabs}
\usepackage{placeins}
\usepackage{xcolor}
\usepackage{hyperref}
\hypersetup{hidelinks}
\usepackage{braket}

\newenvironment{algorithmic}[1][]{%
  \begin{enumerate}\setlength{\itemsep}{1pt}\setlength{\parskip}{0pt}}{%
  \end{enumerate}}
\newcommand{\State}{\item}
\newcommand{\For}[1]{\item \textbf{for} #1 \textbf{do}\begin{enumerate}}
\newcommand{\EndFor}{\end{enumerate}}
\newcommand{\Comment}[1]{\hfill\textit{#1}}

\newcommand{\dt}{\delta t}
\newcommand{\dtau}{\delta\tau}

\newcommand{\chimax}{\chi_{\max}}

\newcommand{\Heff}{H_{\mathrm{eff}}}
\newcommand{\PR}{\mathcal{P}_R}
\newcommand{\PL}{\mathcal{P}_L}
\newcommand{\Xs}{X^{*}}
\newcommand{\sundual}{X^{\odot}}
\newcommand{\Asun}{A^{\odot}}
\newcommand{\Pvw}{P_{V,W}}
\newcommand{\lognorm}{\mu}
\newcommand{\Ree}{\mathrm{Re}}
\DeclareMathOperator{\dist}{dist}
\DeclareMathOperator{\Ran}{Ran}
\DeclareMathOperator{\Ker}{Ker}
\DeclareMathOperator{\spec}{spec}

\begin{document}

\title{Biorthogonal Time-Dependent Variational Principle for Non-Hermitian Systems}

\author{Younes Javanmard}
\email{javanmard.younes@gmail.com}
\affiliation{Institut f\"ur Theoretische Physik, Leibniz Universit\"at Hannover, Appelstra\ss e 2, 30167 Hannover, Germany}

\author{Sina Kazemian}
\email{skazemi5@uwo.ca}
\affiliation{Department of Physics and Astronomy, University of Western Ontario, London, Ontario, Canada}

\date{\today}

\begin{abstract}
We develop a biorthogonal time-dependent variational principle for real-time
dynamics of non-Hermitian quantum many-body systems.  Independent left and
right matrix-product states obey coupled bivariational tangent-space equations
whose cross-Gram matrix defines an oblique projection.  A matrix-free scaled
Taylor action propagates the resulting non-normal local generators without
assembling dense matrices or storing a Krylov basis.  We distinguish the fully
coupled algorithm, which solves the cross-pairing problem and truncates the two
bond bases jointly, from an efficient independently propagated approximation
used for large systems.  Independent truncation can make the retained
left-right pairing nearly singular; overlap drift and the smallest singular
value of the bond cross matrix expose this failure, while coupled truncation
substantially delays it.  Exact benchmarks and convergence tests validate the
method.  Applied to an interacting long-range non-Hermitian Ising chain, it
resolves a biorthogonal dynamical quantum phase transition and shows that a weak
imaginary field shifts the leading critical time from
$t^{\ast}|J|=1.84$ to $1.04$.
\end{abstract}

\maketitle

\section{Introduction}
\label{sec:intro}

Non-Hermitian Hamiltonians describe effective open-system evolution,
gain and loss, asymmetric transport, exceptional points, and the
non-Hermitian skin effect~\cite{Ashida2020,Bergholtz2021}.  Their many-body
dynamics quickly exceed exact-diagonalization limits, motivating
tensor-network methods~\cite{ChenLado2024}.  The central complication is not
only nonunitarity: right and left states are distinct, so the variational
problem is intrinsically paired.

Biorthonormal matrix-product-state (MPS) methods now provide a static
non-Hermitian DMRG framework~\cite{Zhong2025}, but the corresponding dynamical
construction must also control non-normal local generators and the compatibility
of the evolving left and right tangent spaces.  A standard two-site split can
leave each MPS individually well conditioned while making their mutual
cross-pairing nearly singular.  This failure is invisible to either Schmidt
spectrum alone.

Here we provide the dynamical counterpart.  First, we derive the oblique
left-right tangent projection from the time-dependent bivariational
principle~\cite{LowdinMukherjee1972,Arponen1983,PedersenKvaal2019}.  Second, we
adapt a matrix-free scaled Taylor exponential action~\cite{AlMohyHigham2011} to
the local tensor-network generator.  Third, we separate a fully coupled
algorithm---cross-Gram solve plus joint biorthogonal truncation---from a cheaper
paired approximation in which the two MPS are propagated and truncated
independently.  Finally, we connect the loss of compatibility to a measurable
inf--sup constant and show directly that coupled truncation improves both the
cross-pairing condition and the conserved-overlap drift.

The numerical tests combine exact small-system comparisons, bond-dimension and
time-step convergence, and production runs at $L=20,24,32$.  As a physical
application, the paired dynamics resolves a biorthogonal dynamical quantum
phase transition (DQPT) in an interacting long-range Ising chain.  The
functional-analytic details, stochastic Lindblad application, additional
models, and reproducibility data are collected in the Supplemental Material.

\section{Biorthogonal TDVP}
\label{sec:method}

For $H\neq H^{\dagger}$ we propagate an independent ket and dual state,
$|\psi_R(t)\rangle$ under $H$ and $\langle\psi_L(t)|$ under the adjoint
equation.  Their natural observable is
\begin{equation}
  \langle O\rangle(t)=
  \frac{\langle\psi_L(t)|O|\psi_R(t)\rangle}
       {\langle\psi_L(t)|\psi_R(t)\rangle}.
  \label{eq:biorth_obs}
\end{equation}
The initial dual state is part of the initial-value problem; for a physical
initial ket we use $\langle\psi_L(0)|=\langle\psi_R(0)|$.  Biorthogonal
expectation values are generally complex, and the figures show their real parts
unless noted otherwise.  Equation~\eqref{eq:biorth_obs} differs from a
conditional no-jump expectation value, which uses the Hermitian conjugate of a
single propagated ket; that separate application is discussed in the
Supplemental Material.

We impose stationarity of the bivariational action
\begin{equation}
  S[\psi_L,\psi_R]=\int dt\,
  \langle\psi_L(t)|(i\partial_t-H)|\psi_R(t)\rangle
  \label{eq:action}
\end{equation}
on independent left and right MPS manifolds.  Varying their tensors gives
\begin{equation}
  i\,\partial_t|\psi_R\rangle=\PR H|\psi_R\rangle,
  \qquad
  -i\,\partial_t\langle\psi_L|=\langle\psi_L|H\PL,
  \label{eq:projected}
\end{equation}
where $\PR$ and $\PL$ are oblique tangent-space projectors.  In coordinates,
the right update obeys
\begin{equation}
  \sum_j G^{LR}_{ij}\dot\theta_{R,j}=-iF_i,
  \qquad
  F_i=\langle\partial_{L,i}\psi_L|H|\psi_R\rangle,
  \label{eq:tdvp_eom}
\end{equation}
with the cross-Gram matrix
\begin{equation}
  G^{LR}_{ij}=\langle\partial_{L,i}\psi_L|
                       \partial_{R,j}\psi_R\rangle.
  \label{eq:metric}
\end{equation}
An adjoint system propagates the left tensors under $H^{\dagger}$.  Unlike the
Hermitian TDVP metric, $G^{LR}$ need not be Hermitian or positive definite.
Canonical MPS gauges remove the usual parametrization redundancy but do not
control the mutual left-right pairing.

\subsection{Oblique tangent-space geometry: a Banach dual-pair
formulation}
\label{subsec:banach_interpretation}
\label{app:banach}

\paragraph{Dual-pair construction.}
The functional-analytic structure is Petrov--Galerkin rather than orthogonal:
The right tangent manifold supplies trial directions and the left tangent
manifold supplies continuous linear tests.  Let $X$ be a complex Banach space,
$X^*$ its continuous dual, and identify the quantum pair with a primal state
$x\in X$ and an independent dual state $\ell\in X^*$.  With $A=-iH$, the exact
contragredient equations are
\begin{equation}
  \dot x=Ax,
  \qquad
  \dot\ell=-A^*\ell,
  \qquad
  \frac{d}{dt}\ell(x)=0.
  \label{eq:banach_pairing_main}
\end{equation}
This statement uses only the duality pairing $\ell(x)$; no Riesz
identification of $X$ with $X^*$ is required.

At the current state pair, define the gauge-fixed trial and test spaces
$V=T_x\mathcal M_R\subset X$ and $W=T_\ell\mathcal M_L\subset X^*$.  For trial
basis vectors $\{\xi_j\}$ and test functionals $\{\eta_i\}$, introduce
\begin{equation}
  \Xi c=\sum_j c_j\xi_j,
  \qquad
  (Ev)_i=\eta_i(v).
  \label{eq:banach_maps_main}
\end{equation}
Testing the residual $\dot x-Ax$ against $W$ yields the coordinate equation
$G^{LR}\dot{\bm\theta}_R=EAx$, with $G^{LR}_{ij}=\eta_i(\xi_j)$.  Hence
\begin{equation}
  G^{LR}=E\Xi,
  \qquad
  \Pvw=\Xi(E\Xi)^{-1}E.
  \label{eq:banach_oblique_projector}
\end{equation}
For equal-dimensional compatible spaces this is an idempotent projector with
\begin{equation}
  \Ran\Pvw=V,
  \qquad
  \Ker\Pvw=W_\perp.
  \label{eq:banach_projector_geometry_main}
\end{equation}
Here $W_\perp$ is the annihilator of $W$: its vectors are invisible to every
chosen left test, rather than orthogonal in a prescribed inner product.  If
$\dim V>\dim W$, a nonzero trial direction necessarily lies in $W_\perp$; the
restricted pairing is then degenerate.

\paragraph{Stability and MPS compatibility.}
Compatibility is quantified without choosing tangent bases by the inf--sup
constant
\begin{align}
  \beta(V,W)
  &=\inf_{0\neq v\in V}\sup_{0\neq\eta\in W}
    \frac{|\eta(v)|}{\|v\|\,\|\eta\|},
  \label{eq:banach_infsup}\\
  \|\Pvw\|&\leq\beta(V,W)^{-1},
  \label{eq:proj_bound}
  \\
  \|u-\Pvw u\|
  &\leq\left(1+\beta^{-1}\right)\dist(u,V).
  \label{eq:quasi_optimality_main}
\end{align}
The second estimate separates best-approximation error from amplification by
an incompatible test space.  The norms entering $\beta$ are part of the
dual-pair model.  In norm-preserving trial and test coordinates, $\beta$ is the
smallest singular value of $G^{LR}$; otherwise the matrix condition number also
contains coordinate conditioning, which is what the canonical gauge of
Sec.~\ref{sec:method} removes. The measured bond compatibility is then
\begin{equation}
  \beta_b=\sigma_{\min}(M_b),
  \qquad
  (M_b)_{ji}=\langle l_j|r_i\rangle,
  \label{eq:beta_bond}
\end{equation}
where $\{|r_i\rangle\}$ and $\{|l_j\rangle\}$ are the retained right and left
bond bases.  A useful global companion is
\begin{equation}
  \beta(t)=
  \frac{|\langle\psi_L(t)|\psi_R(t)\rangle|}
       {\|\psi_L(t)\|\,\|\psi_R(t)\|}.
  \label{eq:beta_global_main}
\end{equation}
Equation~\eqref{eq:beta_global_main} is the rank-one specialization of
Eq.~\eqref{eq:banach_infsup}, and $\beta^{-2}$ is the Petermann factor of the
pair, so the bound~\eqref{eq:proj_bound} is the excess-noise amplification
familiar from non-Hermitian spectral
projectors~\cite{Petermann1979,Berry2003}.

These local and global quantities diagnose different phenomena.  Gauge
redundancy is a parametrization defect removed by canonicalization; small
$\beta_b$ is a geometric mismatch of the retained trial and test spaces; and
small global $\beta(t)$ is an ill-conditioned state pairing caused by
non-normal norm growth. Only the second is specific to the variational
ansatz: the first is removed by a gauge, the second by coupled truncation, and
the third is a property of the exact trajectory that can limit any biorthogonal
calculation even at full bond dimension. Separate left/right SVDs can drive $M_b$ toward
singularity even while both Schmidt spectra remain benign.  Joint
biorthogonal truncation instead selects the retained bases together and keeps
their restricted pairing compatible.

\paragraph{Conservation and scope.}
Pairing conservation persists under the continuous projected equations when
the manifolds include amplitude variations: testing the primal residual with
$\ell\in W$ and the dual residual with $x\in V$ gives
\begin{equation}
  \frac{d}{dt}\ell(x)=0.
  \label{eq:banach_projected_pairing_main}
\end{equation}
Consequently, discrete overlap drift measures internal inconsistency produced
by splitting, regularization, approximate propagation, MPO compression, and
especially independent truncation.  It is not, by itself, an observable-error
bound; convergence and exact-reference comparisons remain necessary.

The general Banach statement needs two qualifications.  For a generator of
only a forward $C_0$ semigroup, the dual evolution may be merely weak-$*$
continuous and conservation is naturally formulated as a forward--backward
pairing.  Moreover, transient amplification of a non-normal propagator is
controlled by its logarithmic norm rather than by the spectral abscissa alone.
In the finite-dimensional lattices studied here, $A$ is bounded and generates
a group, so the domain issue disappears.  The propagated-error bound, proofs,
crossnorm dependence, semigroup qualifications, and the precise points at
which the MPS implementation re-enters a Hilbert structure are collected in
the Supplemental Material.

\subsection{Practical two-site algorithm}
\label{sec:algorithm}

The continuous equations define the fully coupled target dynamics.  Our coupled
implementation uses the partner MPS as the local test state, solves the
cross-Gram block, and chooses the two retained bond spaces jointly.  For the
large-system production runs we also use a less expensive approximation: the
right MPS is swept under $H$ using its own environments, the left MPS is swept
under $H^{\dagger}$ using its own environments, and the two updated tensors are
split by independent SVDs.  This approximation avoids the cross-pairing solve
but does not exactly realize Eq.~\eqref{eq:projected}; the diagnostics below
measure its departure from the coupled flow.

At a bond $(i,i+1)$ the two-site tensor $\theta$ is advanced by
\begin{equation}
  \theta'=e^{\Delta t\,\Heff}\theta,
  \qquad \Delta t=-i\tau,
  \label{eq:expm}
\end{equation}
where $\Heff$ is applied by contractions of the MPO with the left and right
environments.  The dense local matrix is never assembled.  With
$A=\Delta t\,\Heff$, choose $s$ so that $\|A\|/s\lesssim1$ and apply
\begin{equation}
  \theta_{j+1}\approx
  \sum_{m=0}^{M}\frac{1}{m!}\left(\frac{A}{s}\right)^m\theta_j,
  \qquad j=0,\ldots,s-1,
  \label{eq:scaled_taylor}
\end{equation}
with $\theta_0=\theta$.  Each Taylor term requires one further matrix-free
application of $\Heff$; the sum stops at a relative tolerance and $s$ is raised
if it fails to converge by order $M$.  This stores no Krylov basis and behaves
robustly for the non-normal generators tested here.  Replacing $-i\tau$ by
$-\tau$ gives a formal imaginary-time connection to non-Hermitian DMRG; we do
not benchmark that branch here.

After every real-time update, a two-site tensor is split and truncated to
$\chimax$.  The independent production scheme performs this operation
separately on the two states.  The coupled scheme instead diagonalizes the
cross-density matrix and constructs jointly biorthonormal retained bases, with a
conditioning fallback when the bond pairing approaches singularity.  Full sweep
details and pseudocode are in the Supplemental Material.

\section{Diagnostics and validation}
\label{sec:results}

The exact paired Schr\"odinger evolution conserves
$\mathcal O_{\rm LR}(t)=\langle\psi_L(t)|\psi_R(t)\rangle$.  We reconstruct the
unnormalized overlap from the scale factors removed during canonicalization and
monitor
\begin{equation}
  \epsilon_{\rm ov}(t)=
  \left|\frac{\mathcal O_{\rm LR}(t)}
                   {\mathcal O_{\rm LR}(0)}-1\right|.
  \label{eq:drift}
\end{equation}
The reliable window is determined jointly from overlap drift, discarded weight,
bond-dimension convergence, and time-step convergence.  Crucially,
$\epsilon_{\rm ov}$ diagnoses violation of the paired conservation law; it is
not a bound on the error of an arbitrary observable.

\begin{figure*}[t]
  \begin{minipage}[t]{0.485\textwidth}
    \vspace{0pt}
    \centering
    \includegraphics[width=\linewidth]{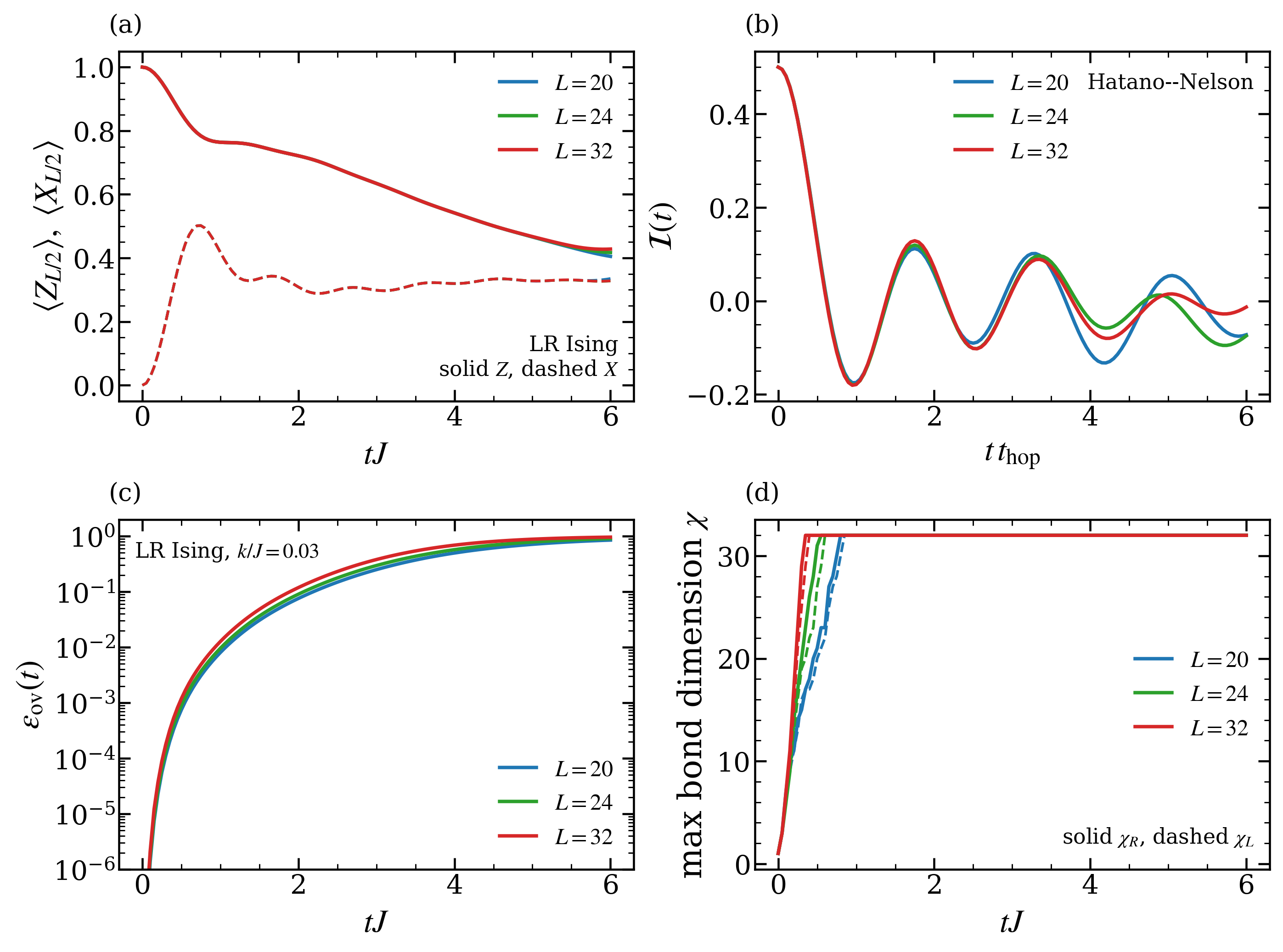}
    \caption{Representative independently propagated MPS trajectories at
    $L=20,24,32$ and $\chi=32$.  (a) Long-range non-Hermitian Ising center-site
    observables at $k/J=0.03$.  (b) Hatano--Nelson imbalance.  (c) Relative
    overlap drift of Eq.~\eqref{eq:drift}; it is a consistency diagnostic, not
    an observable-error bound.  (d) Maximum left (dashed) and right (solid)
    bond dimensions.}
    \label{fig:tdvp_examples}
  \end{minipage}\hfill
  \begin{minipage}[t]{0.485\textwidth}
    \vspace{0pt}
    \centering
    \includegraphics[width=\linewidth]{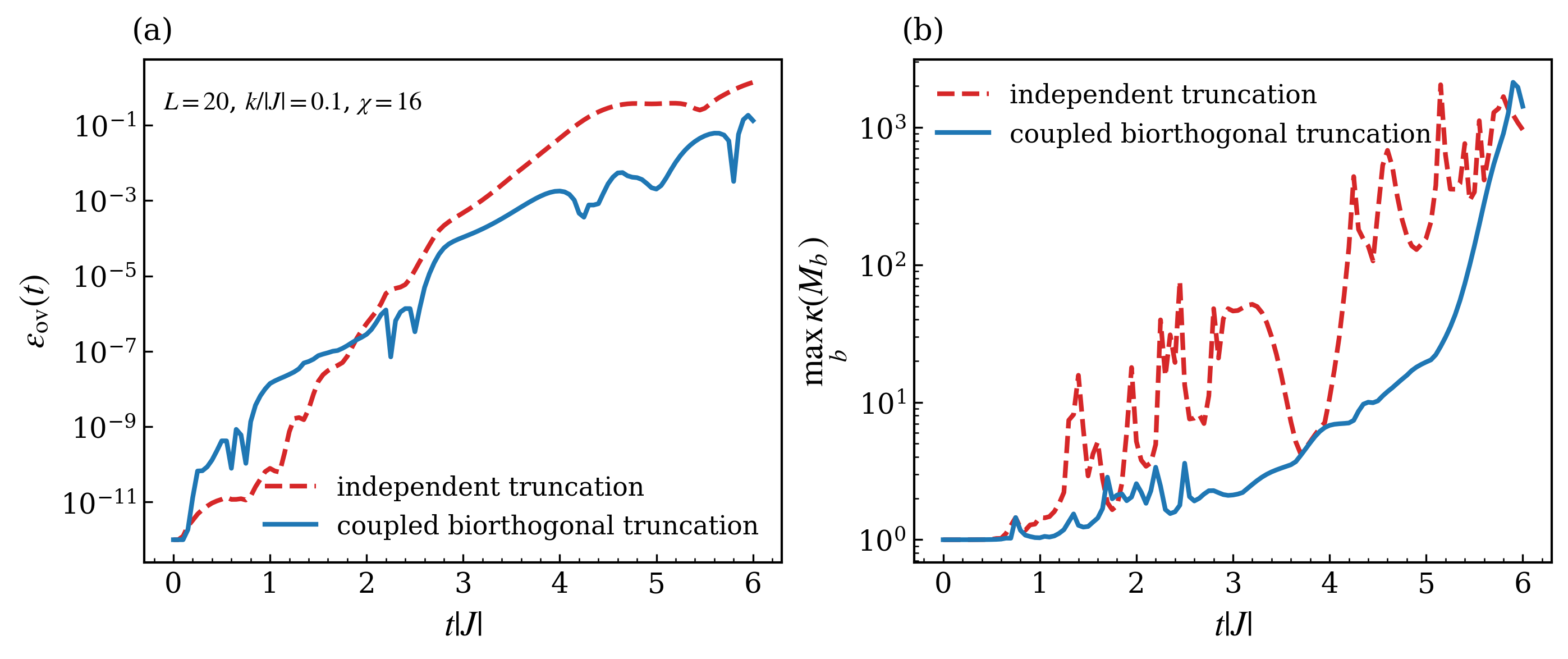}
    \caption{Independent (red dashed) versus coupled biorthogonal truncation
    (blue) for the same long-range non-Hermitian Ising trajectory
    ($L=20$, $k/|J|=0.1$, $\chi=16$).  (a) Relative overlap drift.
    (b) Worst-bond condition number of the retained cross matrix.  Joint
    truncation delays the loss of left-right compatibility.}
    \label{fig:coupled_main}
  \end{minipage}
\end{figure*}

We study a long-range non-Hermitian transverse-field Ising chain and a
disordered Hatano--Nelson chain; Hamiltonians and spin mappings are specified in
the Supplemental Material.  Exact benchmarks give errors below $10^{-11}$ for a
nearest-neighbor non-Hermitian Ising chain, below $10^{-5}$ for the disordered
Hatano--Nelson chain, and below $10^{-5}$ for the long-range Ising observables at
the accessible exact sizes.  Figure~\ref{fig:tdvp_examples} shows the production
regime.  Observables agree across $L=20,24,32$, while the overlap drift and bond
growth expose the finite-entanglement window.  The detailed convergence scan and
the non-Hermiticity dependence of the drift are moved to the Supplemental
Material.

The central stability test is an intervention on the truncation itself.
Figure~\ref{fig:coupled_main} compares independent SVDs with joint
biorthogonal truncation on the same $L=20$ Ising trajectory.  The coupled scheme
holds the overlap drift about an order of magnitude lower through most of the
reliable window and keeps the worst bond condition number near
$\mathcal O(1$--$10)$ substantially longer.  The simultaneous improvement of
$\epsilon_{\rm ov}$ and $\kappa(M_b)$ ties the drift to loss of cross-pairing
compatibility rather than to normalization or a basis artifact.

\section{Biorthogonal dynamical quantum phase transition}
\label{sec:dqpt}

For an interacting non-Hermitian chain, the gauge-invariant biorthogonal return
rate is
\begin{equation}
\begin{aligned}
  \lambda_{\rm bi}(t)&=-\frac{1}{L}\ln|F(t)|,\\
  F(t)&=
  \frac{\langle\psi_L(0)|\psi_R(t)\rangle
        \langle\psi_L(t)|\psi_R(0)\rangle}
       {\langle\psi_L(t)|\psi_R(t)\rangle
        \langle\psi_L(0)|\psi_R(0)\rangle}.
\end{aligned}
\label{eq:rate_function}
\end{equation}
It is invariant under independent complex rescalings of either state and
reduces to the ordinary Loschmidt echo in the Hermitian limit.  A DQPT is a zero
of a return amplitude, not of the conserved equal-time pairing.  Both branches
are therefore essential; the extended forward/backward interpretation and the
free-fermion limit are given in the Supplemental Material.

\begin{figure*}[t]
  \centering
  \includegraphics[width=0.98\textwidth]{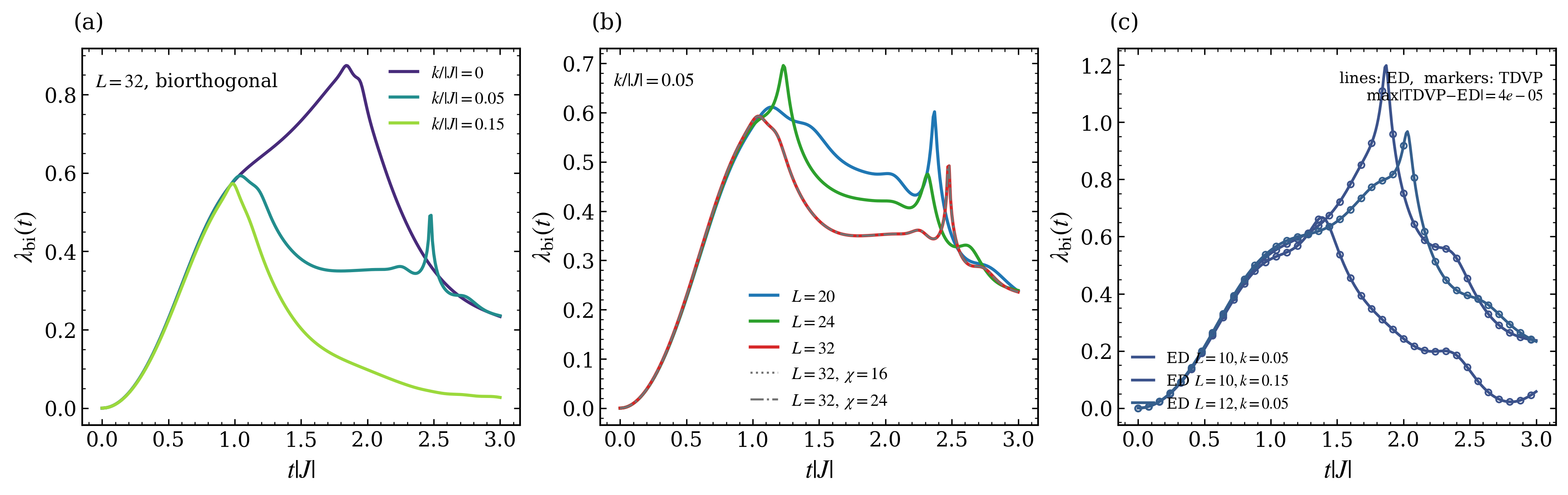}
  \caption{Biorthogonal DQPT in the long-range non-Hermitian Ising chain.
  (a) Dependence on the imaginary field at $L=32$: the leading cusp shifts from
  $t^{\ast}|J|=1.84$ to $1.04$ and broadens at stronger non-Hermiticity.
  (b) Size and bond-dimension convergence at $k/|J|=0.05$.
  (c) Independent paired-TDVP validation against exact biorthogonal evolution.}
  \label{fig:dqpt}
\end{figure*}

We quench $|+\rangle^{\otimes L}$ into a ferromagnetic long-range
non-Hermitian Ising chain with $J=-1$, $h=0.5$, and $\alpha=6$.
Figure~\ref{fig:dqpt} shows that the Hermitian cusp in
$t^{\ast}|J|=1.84$ moves to $1.04$ in $k/|J|=0.05$.  At $k/|J|=0.15$ the sharp
cusp is replaced by a broad maximum at accessible sizes; without a
finite-size analysis at this coupling, we do not infer the absence of a
thermodynamic nonanalyticity.  At $k/|J|=0.05$ the leading cusp is stable across
$L=20,24,32$ and under $\chi=16,24,32$ refinement.  Independent-propagation
results agree with exact biorthogonal evolution to $4\times10^{-5}$ at
$L=10,12$.

\begin{figure*}[t]
  \centering
  \includegraphics[width=0.86\textwidth]{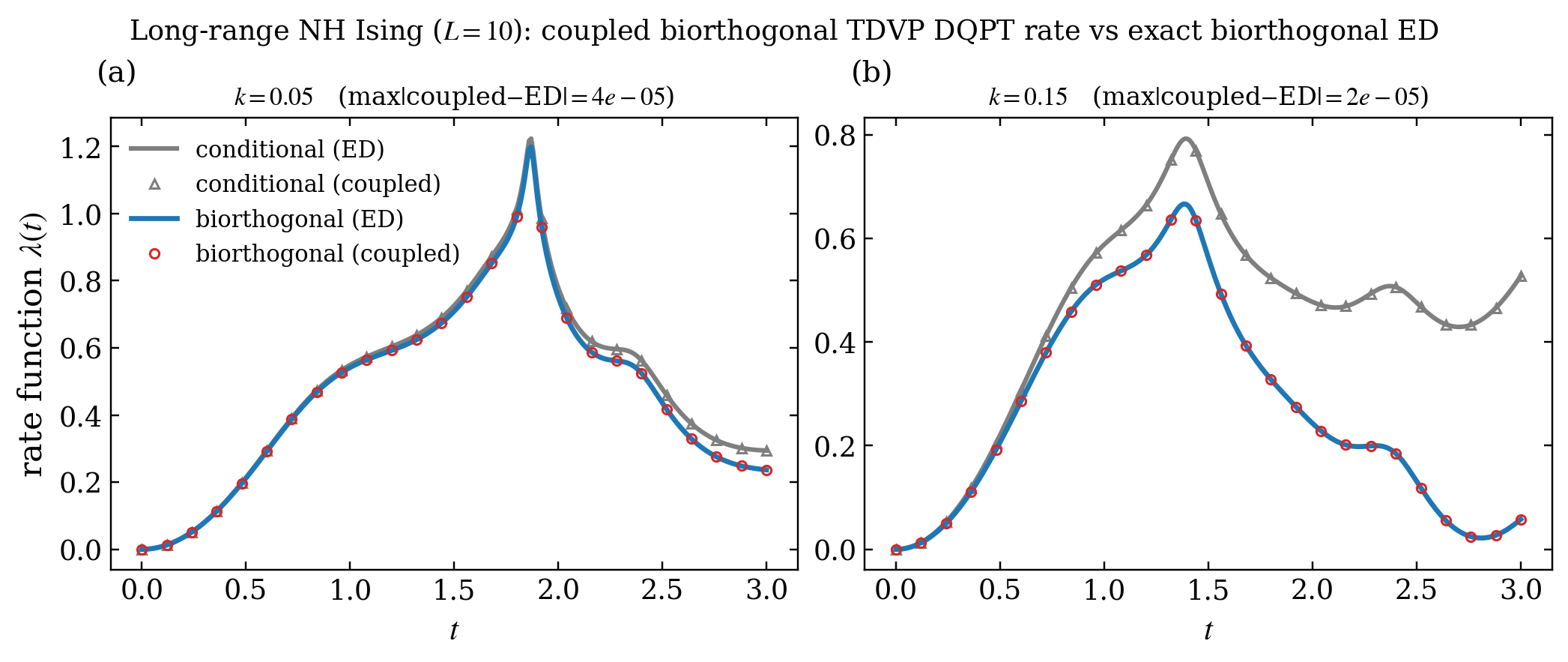}
  \caption{Fully coupled oblique TDVP for the DQPT rate at $L=10$.
  The coupled calculation (markers) reproduces exact biorthogonal ED (lines)
  through the cusp to $2$--$4\times10^{-5}$ at
  (a) $k/|J|=0.05$ and (b) $0.15$.  The conditional single-branch rate (gray)
  differs from the biorthogonal rate and is shown for comparison only.}
  \label{fig:dqpt_coupled_main}
\end{figure*}

Most importantly, the direct-coupled update yields the same result.  Solving the
cross-Gram problem with the partner tangent space reproduces exact
biorthogonal diagonalization through the cusp to $2$--$4\times10^{-5}$ at
$L=10$ (Fig.~\ref{fig:dqpt_coupled_main}).  The observed transition is therefore
not an artifact of the independent production approximation.

\section{Conclusion}
\label{sec:conclusion}

We formulated non-Hermitian MPS dynamics as a paired bivariational TDVP with an
oblique left-right tangent projection and implemented its local evolution with a
matrix-free scaled Taylor action.  This avoids dense local generators and makes
the role of the cross-Gram compatibility explicit.

Exact benchmarks, convergence tests, and the coupled-truncation intervention
show both where the efficient independent approximation works and how it fails.
Joint truncation delays the collapse of the retained cross-pairing, while
overlap drift provides a useful internal warning without being promoted to an
observable-error bound.  The DQPT calculation then demonstrates the method on an
interacting system and resolves a non-Hermiticity-induced shift of the leading
critical time.

The principal limitation is conditioning near a biorthogonal singularity.  The
fully coupled algorithm controls this more faithfully but adds a non-Hermitian
cross-pairing solve at every bond; the independent scheme remains the practical
large-system approximation when its diagnostics stay converged.  Developing
more efficient structure-preserving truncations is the natural next step,
for which the Supplemental Material supplies a cautionary constraint: truncating
against the conserved functional alone is not the answer, since it preserves the
invariant exactly while discarding the dominant Schmidt weight
(Sec.~\ref{app:ordertrunc}).

\paragraph{Data availability}The data and source code generated and/or analyzed during the current study are available from the corresponding author upon reasonable request.

\bibliography{paper}

\clearpage
\onecolumngrid
\section*{Supplemental Material}
\suppressfloats[t]
\setcounter{equation}{0}
\renewcommand{\theequation}{S\arabic{equation}}
\renewcommand{\theHequation}{S\arabic{equation}}
\setcounter{section}{0}
\renewcommand{\thesection}{S\arabic{section}}
\renewcommand{\theHsection}{S\arabic{section}}
\setcounter{figure}{0}
\renewcommand{\thefigure}{S\arabic{figure}}
\renewcommand{\theHfigure}{S\arabic{figure}}
\setcounter{table}{0}
\renewcommand{\thetable}{S\arabic{table}}
\renewcommand{\theHtable}{S\arabic{table}}

\section{Stochastic Lindblad Unraveling}
\label{sec:lindblad_mc}

The matrix-free local exponential also applies to conditional quantum
trajectories.  This use is distinct from the biorthogonal construction in the
main text: a no-jump trajectory contains one ket
$|\psi(t)\rangle=e^{-iH_{\rm eff}t}|\psi(0)\rangle$, and its observables use the
Hermitian-conjugate bra,
\begin{equation}
  \langle O\rangle_{\rm cond}=
  \frac{\langle\psi(t)|O|\psi(t)\rangle}
       {\langle\psi(t)|\psi(t)\rangle}.
  \label{eq:cond_obs}
\end{equation}
For a Markovian open system,
\begin{equation}
  \dot\rho=-i[H,\rho]+\sum_\mu\left(
  L_\mu\rho L_\mu^\dagger-
  \frac{1}{2}\{L_\mu^\dagger L_\mu,\rho\}\right),
  \label{eq:lindblad}
\end{equation}
the Monte Carlo wave-function method~\cite{Dalibard1992,Molmer1993,Plenio1998}
propagates between jumps with
$H_{\rm eff}=H-\frac{i}{2}\sum_\mu L_\mu^\dagger L_\mu$ and samples channel
$\mu$ with probability
$p_\mu=\delta t\,\langle L_\mu^\dagger L_\mu\rangle+O(\delta t^2)$.

\begin{figure}[!b]
  \centering
  \includegraphics[width=0.96\columnwidth]{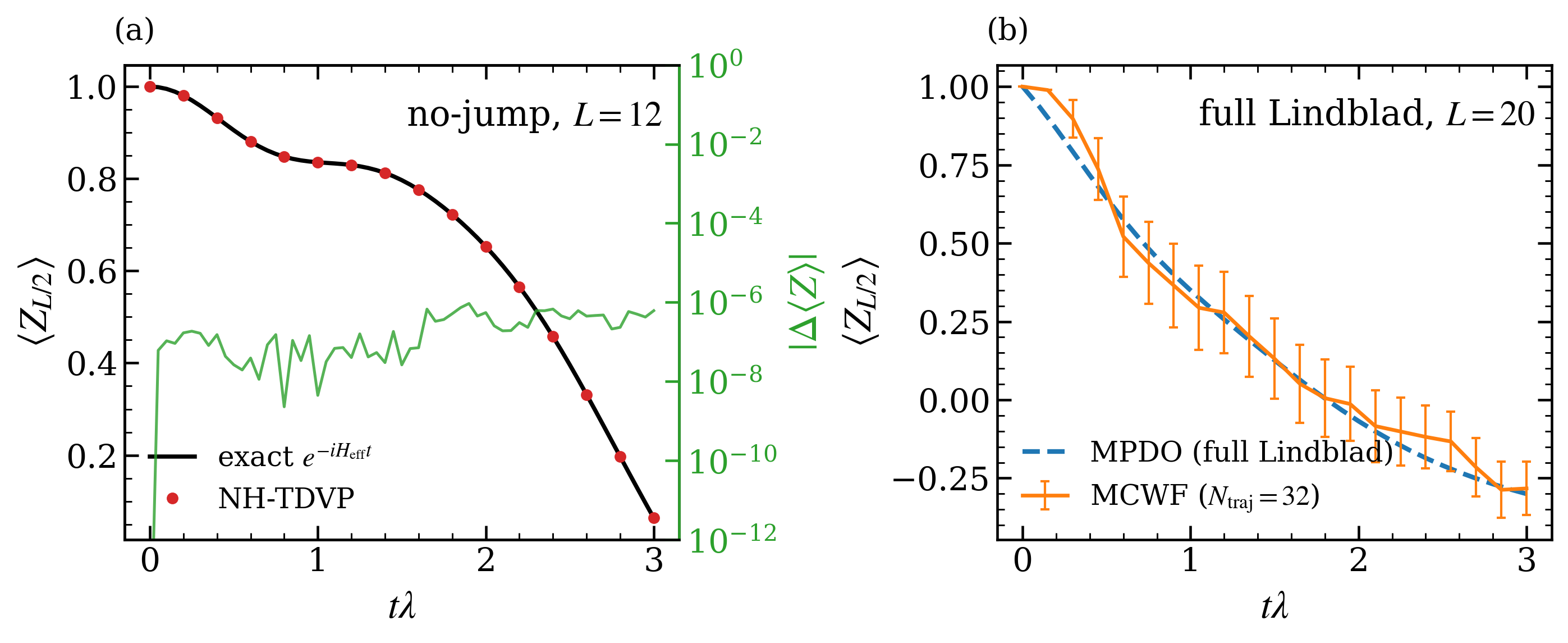}
  \caption{Amplitude-damped transverse-field Ising chain
  ($\lambda=1$, $h=0.5$, $\gamma=0.3$).  (a) No-jump TDVP (markers) against exact
  evolution (line) at $L=12$; the absolute difference stays
  $\lesssim10^{-6}$.  (b) Quantum-jump trajectory average at $L=20$ (markers)
  against an independent MPDO calculation (dashed), with statistical error
  bars.}
  \label{fig:lindblad}
\end{figure}

For amplitude damping, $L_i=\sqrt{\gamma}\,\sigma_i^-$, the no-jump generator is
\begin{align}
  H_{\rm eff}
  &= -\lambda\sum_i Z_iZ_{i+1}-h\sum_i X_i
     -\frac{i\gamma}{2}\sum_i n_i \notag\\
  &= -\lambda\sum_i Z_iZ_{i+1}-h\sum_i X_i
     -\frac{i\gamma}{4}\sum_i Z_i+{\rm const.}
  \label{eq:amp_damp_heff}
\end{align}
The TDVP sweep supplies the no-jump propagation; stochastic trajectory averages
then reconstruct the Lindblad expectation values.  Figure~\ref{fig:lindblad}
validates the deterministic branch against exact evolution at $L=12$ and the
full trajectory average against an independent matrix-product density-operator
simulation at $L=20$~\cite{MPDO2022}.  This is a validation of the propagator,
not of the paired left-right geometry.

\section{Banach Dual-Pair Formulation: Functional-Analytic Supplement}
\label{app:banach_analysis}

This section collects four technical points underlying the dual-pair formulation
of Sec.~\ref{app:banach}. Sections~\ref{app:sundual} and~\ref{app:fwdbwd} concern
the regularity and the correct form of the pairing conservation law when the
generator is only a forward semigroup; Sec.~\ref{app:lognorm} identifies the
constant that governs transient growth for a non-normal generator, and connects
it to the local propagator; and Sec.~\ref{app:hilbertian} separates the three
distinct places at which the matrix-product implementation reverts to a Hilbert
structure, one of which is removable by a change of parametrization. None of these is required to follow the algorithm, but each is needed
for the continuous statements of Sec.~\ref{app:banach} to be exact rather than
formal.

Let $A:D(A)\subset X\to X$ be a densely defined generator and $A^*$ its Banach
dual.  The exact contragredient evolution is
\begin{equation}
  \dot x=Ax,
  \qquad
  \dot\ell=-A^*\ell,
  \label{eq:banach_pair_evolution}
\end{equation}
and, whenever both trajectories are differentiable on the same interval,
\begin{equation}
  \frac{d}{dt}\ell(x)=\dot\ell(x)+\ell(\dot x)
  =-(A^*\ell)(x)+\ell(Ax)=0.
  \label{eq:banach_pairing_conservation}
\end{equation}
For the projected evolution, the inf--sup bound in the main text separates
modelling error from test-space amplification.  If $x$ solves $\dot x=Ax$,
$x_h$ solves $\dot x_h=P_{V(t),W(t)}Ax_h$, and
$\|T(t)\|\leq Me^{\omega t}$, then
\begin{equation}
\begin{aligned}
  \|x(t)-x_h(t)\|
  &\leq\left(1+\frac{1}{\beta_*}\right)M\\[-1mm]
  &\quad\times\int_0^t e^{\omega(t-s)}
  \dist\!\left(Ax_h(s),V(s)\right)\,ds,\\
  \beta_*&=\inf_{s\leq t}\beta(V(s),W(s)).
\end{aligned}
  \label{eq:projection_error_bound}
\end{equation}
The following subsections state the semigroup qualifications and identify the
appropriate non-normal growth constant.

\begin{table}[t]
  \centering
  \footnotesize
  \renewcommand{\arraystretch}{1.12}
  \caption{Correspondence between the biorthogonal quantum formulation and the
  primal--dual Banach-space formulation.}
  \label{tab:banach_dictionary}
  \resizebox{0.96\columnwidth}{!}{%
  \begin{tabular}{@{}lll@{}}
    \toprule
    Biorthogonal notation & Dual-pair notation & Meaning \\
    \midrule
    $|\psi_R\rangle$ & $x\in X$ & primal state \\
    $\langle\psi_L|$ & $\ell\in X^*$ & independent dual state \\
    $-iH$ & $A:X\to X$ & primal generator \\
    adjoint evolution & $-A^*:X^*\to X^*$ & dual generator \\
    $\langle\psi_L|\psi_R\rangle$ & $\ell(x)$ & conserved pairing \\
    right tangent space & $V=T_x\mathcal M_R$ & trial directions \\
    left tangent space & $W=T_\ell\mathcal M_L$ & test directions \\
    $G^{LR}_{ij}$ & $\eta_i(\xi_j)$ & restricted pairing \\
    tangent projection & $\Xi(E\Xi)^{-1}E$ & oblique projector \\
    compatibility & $\beta(V,W)$ & inf--sup constant \\
    \bottomrule
  \end{tabular}%
  }
\end{table}

\subsection{Primal--dual Petrov--Galerkin construction}

Let $X$ be a complex Banach space, $X^*$ its continuous dual, and
$\ell(x)=\langle\ell,x\rangle_{X^*,X}$ the duality pairing.  In the quantum
specialization, $x$ and $\ell$ represent $|\psi_R\rangle$ and
$\langle\psi_L|$, respectively.  No Riesz identification is required at the
continuous level; the two trajectories are independent.

At the current pair, define the gauge-fixed tangent spaces
\begin{equation}
  V=T_x\mathcal M_R\subset X,
  \qquad
  W=T_\ell\mathcal M_L\subset X^*.
  \label{eq:banach_tangent_spaces}
\end{equation}
For bases $\{\xi_j\}\subset V$ and $\{\eta_i\}\subset W$, introduce
\begin{equation}
  \Xi c=\sum_j c_j\xi_j,
  \qquad
  (Ev)_i=\eta_i(v).
  \label{eq:banach_trial_test_maps}
\end{equation}
Testing the right residual $r_R=\dot x-Ax$ against every $\eta_i$ gives
\begin{equation}
  G^{LR}\dot{\bm\theta}_R=EAx,
  \qquad
  G^{LR}=E\Xi,
  \qquad
  G^{LR}_{ij}=\eta_i(\xi_j).
  \label{eq:banach_cross_gram}
\end{equation}
For equal-dimensional compatible spaces, $G^{LR}$ is invertible and the
projector in the main text obeys
\begin{equation}
  P_{V,W}^2=P_{V,W},
  \qquad
  \Ran P_{V,W}=V,
  \qquad
  \Ker P_{V,W}=W_\perp,
  \label{eq:banach_projector_properties}
\end{equation}
where
\begin{equation}
  W_\perp=\{v\in X:\eta(v)=0\ \text{for all }\eta\in W\}.
  \label{eq:banach_annihilator}
\end{equation}
Thus the rejected directions are invisible to the test space rather than
orthogonal in a prescribed inner product.  If $\dim V>\dim W$, some nonzero
trial direction lies in $W_\perp$ and $\beta(V,W)=0$; unequal retained
dimensions therefore signal geometric degeneracy rather than merely a
rectangular linear system.

\subsection{Compatibility, quasi-optimality, and pairing conservation}

The inf--sup constant of Eq.~\eqref{eq:banach_infsup} is basis independent but
norm dependent.  In norm-preserving trial and test coordinates it equals the
smallest singular value of $G^{LR}$; in arbitrary coordinates the matrix
condition number also contains parametrization conditioning.  Canonical MPS
gauges are therefore essential when $\sigma_{\min}(M_b)$ is interpreted as a
geometric diagnostic.

For $u\in X$, put $v=P_{V,W}u$.  Since $\eta(v)=\eta(u)$ for all $\eta\in W$,
\begin{equation}
  \beta\|v\|
  \leq\sup_{0\neq\eta\in W}\frac{|\eta(v)|}{\|\eta\|}
  =\sup_{0\neq\eta\in W}\frac{|\eta(u)|}{\|\eta\|}
  \leq\|u\|,
  \label{eq:proj_bound_proof}
\end{equation}
which proves $\|P_{V,W}\|\leq1/\beta$.  Because $P_{V,W}w=w$ for $w\in V$,
the same estimate gives
\begin{equation}
  \|u-P_{V,W}u\|
  \leq\left(1+\frac{1}{\beta}\right)\dist(u,V).
  \label{eq:quasi_optimality}
\end{equation}
In a Hilbert space the Kato--Szyld identity improves the constant to
$1/\beta$~\cite{IpsenMeyer1995,Szyld2006}; the general Banach estimate leads to
Eq.~\eqref{eq:projection_error_bound}.

The norms in $\beta$ are part of the definition.  For finite bond dimension,
norm equivalence bounds the change between the Hilbert crossnorm and admissible
injective or projective crossnorms by
\begin{equation}
  \chi^{-1/2}\leq
  \frac{\beta^{(\pi)}}{\beta^{(2)}},\,
  \frac{\beta^{(\varepsilon)}}{\beta^{(2)}}
  \leq\chi^{1/2},
  \label{eq:crossnorm_bound}
\end{equation}
a factor of $5.7$ at $\chi_{\max}=32$.  The ambiguity becomes unbounded only in
the infinite-dimensional limit, where these crossnorms need not be
equivalent~\cite{Ryan2002}.

Pairing conservation also survives compatible projection.  If the manifolds
admit amplitude variations, so that $x\in V$ and $\ell\in W$, and the dual
residual satisfies
\begin{equation}
  (\dot\ell+A^*\ell)(v)=0
  \quad\text{for all }v\in V,
  \label{eq:banach_dual_pg_condition}
\end{equation}
then testing the primal residual with $\ell$ and the dual residual with $x$
yields
\begin{equation}
  \frac{d}{dt}\ell(x)=0.
  \label{eq:banach_projected_conservation}
\end{equation}
Discrete overlap drift can therefore arise from projector splitting,
regularization, approximate propagation, MPO compression, and especially
independent left/right truncation.

Three distinct degeneracies should not be conflated.  Gauge redundancy is a
parametrization defect removed by canonicalization.  A small inf--sup constant is
a geometric incompatibility between the retained trial and test spaces.  A
small global $\beta(t)$ is instead an ill-conditioned exact state pairing caused
by non-normal norm growth.  The first is removable by gauge, the second is
addressed by coupled truncation, and the third can limit any biorthogonal
calculation even at full bond dimension.

\subsection{Regularity of the dual evolution}
\label{app:sundual}

Equation~\eqref{eq:banach_pair_evolution} presumes that the dual trajectory is
differentiable in the norm of $\Xs$, which is not automatic.  If $A$ generates a
$C_0$ semigroup $T(t)$ on $X$, the adjoint family $T(t)^*$ on $\Xs$ is
weak-$*$ continuous but in general \emph{not} strongly continuous, and $D(A^*)$
need not be norm dense in $\Xs$.  The maximal subspace of strong continuity is
the \emph{sun dual}~\cite{Phillips1955,vanNeerven1992,EngelNagel2000}
\begin{equation}
  \sundual=\overline{D(A^*)}^{\,\|\cdot\|_{\Xs}}\subseteq\Xs ,
  \label{eq:sundual}
\end{equation}
which is $T(t)^*$-invariant and on which $T(t)^*|_{\sundual}$ is a $C_0$
semigroup with generator $\Asun$, the part of $A^*$ in $\sundual$.  Accordingly,
Eq.~\eqref{eq:banach_pair_evolution} holds in the norm sense on $\sundual$ with
$A^*$ replaced by $\Asun$, and on all of $\Xs$ it must be read weak-$*$ly,
\begin{equation}
  \frac{d}{dt}\,\ell(t)(x)=-\ell(t)(Ax)
  \qquad\text{for all }x\in D(A).
  \label{eq:weakstar_dual}
\end{equation}

This is not an academic caveat for the ordered examples discussed below,
since both dual pairs are non-reflexive.  For $X=L^1$ with $\Xs=L^\infty$ the
adjoint (Koopman) semigroup is generically not strongly continuous---for the heat
semigroup on $L^1(\mathbb{R})$ one finds $\sundual=\mathrm{BUC}(\mathbb{R})
\subsetneq L^\infty(\mathbb{R})$---and for
$\mathcal{S}_1(\mathcal{H})^*=\mathcal{B}(\mathcal{H})$ the Heisenberg semigroup
is $\sigma$-weakly continuous, which is precisely why quantum dynamical
semigroups are defined to be normal rather than norm
continuous~\cite{BratteliRobinson1987}.  When $X$ is reflexive,
$\sundual=\Xs$ and the distinction disappears; in the finite-dimensional
many-body setting simulated in this work $A=-iH$ is bounded, $T(t)$ is a group,
and all of these subtleties are vacuous.

\subsection{Forward--backward form of the conservation law}
\label{app:fwdbwd}

Equation~\eqref{eq:banach_pairing_conservation} also presumes that both
trajectories solve initial-value problems on a common interval.  If $A$ generates
only a forward semigroup, the contragredient flow $(T(t)^*)^{-1}$ does not exist
and this is unavailable.  The conservation law nonetheless survives in a
terminal-data form requiring neither invertibility nor strong continuity of the
dual semigroup.  Fix a horizon $T>0$, propagate the primal state forward from
$x_0$ and the dual functional \emph{backward} from terminal data $\ell_T\in\Xs$,
\begin{equation}
  x(t)=T(t)x_0,
  \qquad
  \ell(t):=T(T-t)^*\ell_T,
  \qquad t\in[0,T].
  \label{eq:fwd_bwd_pair}
\end{equation}
Then, for every $t\in[0,T]$,
\begin{equation}
  \ell(t)\bigl(x(t)\bigr)
  =\ell_T\bigl(T(T-t)T(t)x_0\bigr)
  =\ell_T\bigl(T(T)x_0\bigr),
  \label{eq:fwd_bwd_conservation}
\end{equation}
independently of $t$.  This is the pairing conservation in its structurally
correct form: an adjoint-state (forward--backward) duality on a fixed interval,
as in optimal control and in the Schr\"odinger--Heisenberg duality of open-system
dynamics, rather than a pair of simultaneous initial-value problems.  When $T(t)$
extends to a group---the case relevant to the simulations here, where $H$ is a
bounded operator on a finite lattice---one may set $\ell(t)=T(-t)^*\ell_0$ and
recover Eqs.~\eqref{eq:banach_pair_evolution}--%
\eqref{eq:banach_pairing_conservation} as a genuine initial-value pair, with
Eq.~\eqref{eq:banach_pairing_conservation} holding in the norm sense.

\subsection{Non-normality enters through the logarithmic norm}
\label{app:lognorm}

The growth constant $\omega$ in Eq.~\eqref{eq:projection_error_bound} should not
be taken to be $\|A\|$.  The sharp short-time exponential rate in a Banach space
is the \emph{logarithmic norm}, or numerical abscissa,
\begin{equation}
  \lognorm(A)=\lim_{h\to0^{+}}\frac{\|I+hA\|-1}{h},
  \qquad
  \|e^{tA}\|\le e^{\lognorm(A)t},
  \label{eq:log_norm}
\end{equation}
which is well-defined without an inner product and is characterized by the
Lumer--Phillips theorem through dissipativity with respect to a semi-inner
product compatible with $\|\cdot\|$~\cite{LumerPhillips1961,Soderlind2006}.  For
normal generators $\lognorm(A)=\max\Ree\,\spec(A)$; for non-normal generators
$\lognorm(A)$ can exceed the spectral abscissa, and the gap between the two is
exactly the transient amplification that makes Krylov iterations and
imaginary-time filtering fragile in the non-Hermitian setting.  The same
information is encoded spectrally by Banach-space pseudospectra
$\sigma_\varepsilon(A)=\{z:\|(z-A)^{-1}\|>\varepsilon^{-1}\}$, whose excursion
into the right half-plane bounds transient growth from below through the Kreiss
constant~\cite{TrefethenEmbree2005}.

This supplies the constant that the two halves of this work share.  The local
generator $\Heff$ of Eq.~\eqref{eq:expm} is non-normal, and $\lognorm(\dt\Heff)$
governs both the growth factor $e^{\omega t}$ in
Eq.~\eqref{eq:projection_error_bound} and the accuracy of the scaled-Taylor local
propagator, whose substep count is chosen in Algorithm~S1 from the cruder estimate
$\|\dt\Heff\|/s\lesssim1$.  Replacing that estimate by $\lognorm$, or by the
sharper sequence $\|A^m\|^{1/m}$ used in the backward-error analysis of
Ref.~\cite{AlMohyHigham2011}, is the natural refinement and would furnish a
certified local error bound in place of the empirical tolerance test.

\subsection{Where the implementation returns to the Hilbert setting}
\label{app:hilbertian}

The dual-pair geometry of Sec.~\ref{app:banach} requires no inner product, but
the MPS realization reintroduces one at three distinct points, of which the last
two are matters of norm and the first only of convention.

\emph{(a) Parametrization of the dual manifold.}  $\mathcal{M}_L$ is nominally a
submanifold of $\Xs$, yet we do not parametrize functionals directly: we
parametrize a vector by tensors $A_L\in X$ and define
$\ell=\langle\psi_L|\cdot\,\rangle$ through the inner product.  $\mathcal{M}_L$
is therefore the Riesz image of a manifold in $X$, not an intrinsic low-rank
family in $\Xs$, and the map from tensors to functionals is antilinear.  Unlike
(b) and (c) this obstruction is removable without leaving the present setting:
the strict bivariational convention of Refs.~\onlinecite{LowdinMukherjee1972}
and~\onlinecite{Arponen1983} stores the components of $\ell$ itself, so that the
parametrization is linear, the action is jointly holomorphic in
$(\bm\theta_L,\bm\theta_R)$, and $\mathcal{M}_L$ becomes a low-rank format
\emph{for functionals}.  The two conventions differ by a relabelling of the
stored left tensors---the environment contractions of Sec.~\ref{app:coupled_update}
then carry $B^{[i]}$ in place of $B^{[i]*}$---and we retain the conjugating one
because it matches the standard tensor-network canonical forms; a genuinely
Banach construction would require the linear one, since no Riesz map exists.

\emph{(b) The tensor-product norm.}  Writing the many-body space as
$X=\bigotimes_{i=1}^{L}X_i$ is unambiguous only in the Hilbert case.  On Banach
spaces the algebraic tensor product admits a continuum of reasonable crossnorms
between the injective $\|\cdot\|_\varepsilon$ and the projective
$\|\cdot\|_\pi$, inequivalent in infinite dimensions, giving different
completions and different duals; $(\bigotimes_\pi X_i)^*$ is the space of bounded
$L$-linear forms, which is where a dual ``MPS'' would actually
live~\cite{Ryan2002}.  The matrix-product format, and every statement about bond
dimension, is relative to that choice.

\emph{(c) Truncation.}  There is no Eckart--Young theorem outside Hilbert space.
Existence of a best rank-$\chi$ approximation in a Banach tensor format requires
additional hypotheses---weak closedness of the format together with
reflexivity-type conditions on the crossnorm---and can fail in exactly the spaces
invoked by those examples, since $L^1$ and $L^\infty$ are
not reflexive~\cite{Hackbusch2012,FalcoHackbusch2012}.  Quasi-optimal truncation
must then be obtained from a norm-adapted factorization, and the truncation error
can no longer be read off a discarded singular-value weight.

The claim of Sec.~\ref{app:banach} is therefore that the \emph{continuous
variational geometry} is Banach; the discrete algorithm is not, and making it so
is a problem in numerical tensor calculus rather than a change of notation.

\subsection{Order-adapted truncation on an ordered dual pair}
\label{app:ordertrunc}

An ordered dual pair supplies the positivity and normalization absent from a
generic Banach pair.  A base-norm space $X$ carries a positive cone and a base
selected by a strictly positive functional; its dual $X^*$ is an order-unit
space with unit $e$.  For a normalized positive $x$, base-norm and order-unit
normalization give
\begin{equation}
  \beta=\frac{e(x)}{\|e\|_{X^*}\,\|x\|_X}=1.
  \label{eq:cone_beta}
\end{equation}
This applies to Markov evolution on $L^1/L^\infty$ and to completely positive
trace-preserving evolution on the trace-class/bounded-operator pair, with total
mass or trace as $e$. Equation~\eqref{eq:cone_beta} is definitional rather
than derived---all three quantities in it are fixed by the definition of an
ordered pair---and that is precisely its content: what an ordered pair supplies
is a normalization, not a dynamical property.  The statement is specific to the
normalization functional and does not prevent higher-dimensional tangent-space
incompatibility.  It also does not apply to the biorthogonal state pairing of
the non-Hermitian wave-function problem, which has no corresponding positive
order unit---which is why that problem needs the measured pairing
diagnostics of Sec.~\ref{app:crosspairing} in place of a structural guarantee.

The remainder of this section tests the truncation obstruction of
Sec.~\ref{app:hilbertian}(c) numerically, in a setting already present in this
work, and the result corrects the expectation that motivates it.

The Lindblad simulation of Sec.~\ref{sec:lindblad_mc} is an instance of this
ordered dual pair: $X=\mathcal{S}_1(\mathcal{H})$ is a
base-norm space whose base is fixed by the trace, $\Xs=\mathcal{B}(\mathcal{H})$
is an order-unit space with $e=\mathbf{1}$, and Eq.~\eqref{eq:lindblad} generates
a positive trace-preserving semigroup, for which the rank-one compatibility
constant is pinned by Eq.~\eqref{eq:cone_beta}.  The matrix-product
density-operator representation of $\rho(t)$ is nevertheless truncated by singular
value decomposition in the $2$-norm, which is \emph{not} the base norm.  The
mismatch that Sec.~\ref{app:hilbertian}(c) describes in the abstract is therefore
already present in a calculation we perform, and can be measured rather than
argued.

We compare two truncations of the same amplitude-damped MPDO trajectory
($L=20$, $\gamma=0.3$) against a $\chi_{\rm ref}=48$ reference: (i) ordinary
singular value truncation, retaining the $\chi$ largest singular values, and
(ii) an \emph{order-adapted} truncation in which the retained bond subspace is
chosen by its pairing with the order unit---the
$\mathcal{S}_1/\mathcal{B}(\mathcal{H})$ instance of the cross-density-matrix
construction of Sec.~\ref{app:coupled}---so that $\operatorname{Tr}\rho$ is
preserved by construction.  Both schemes are applied to the same evolution, so
the comparison isolates the truncation operator.

\begin{figure*}[t]
  \centering
  \includegraphics[width=0.98\textwidth]{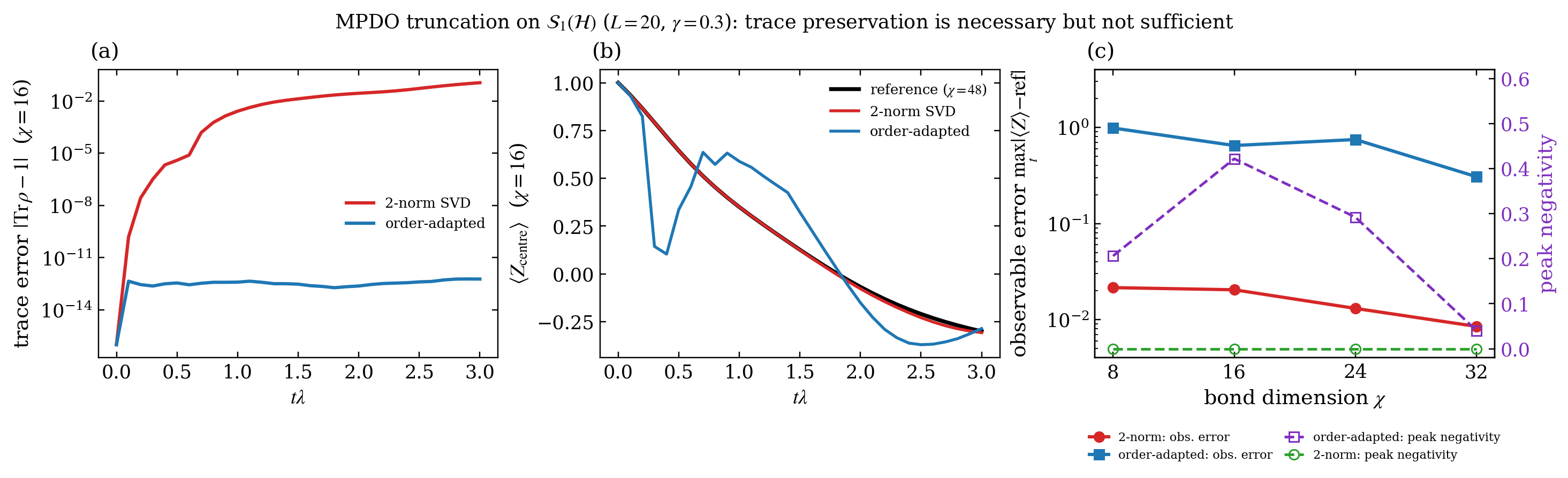}
  \caption{Order-adapted versus $2$-norm truncation of the matrix-product density
  operator on the ordered dual pair $\mathcal{S}_1(\mathcal{H})$, for the
  amplitude-damped chain of Sec.~\ref{sec:lindblad_mc} ($L=20$, $\gamma=0.3$),
  against a $\chi_{\rm ref}=48$ reference.  (a)~Trace error
  $|\operatorname{Tr}\rho-1|$ at $\chi=16$: the order-adapted scheme preserves the
  base functional by construction, while $2$-norm truncation leaks to
  $\sim10^{-2}$.  (b)~Center-site $\langle Z\rangle$ at $\chi=16$: the $2$-norm
  trajectory lies on the reference to line width, while the trace-preserving one
  does not track the dynamics.  (c)~Observable error $\max_t|\langle Z\rangle-{\rm
  ref}|$ (left axis) and peak local negativity (right axis) versus $\chi$.
  Preserving the pairing exactly is not sufficient: selecting components by their
  pairing with the order unit is insensitive to the weight they carry.}
  \label{fig:ordertrunc}
\end{figure*}

Figure~\ref{fig:ordertrunc} reports the outcome, and it is not the one the
abstract argument suggests.  Panel~(a) is as expected: the order-adapted scheme
holds the trace at $\sim10^{-13}$ across the window while $2$-norm truncation
leaks to $\sim10^{-2}$.  This much is definitional---retaining the subspace that
carries the trace preserves the trace exactly---and it confirms that $2$-norm
truncation does violate the invariant that Eq.~\eqref{eq:cone_beta} protects.
Panel~(b) shows the price.  The order-adapted trajectory fails to track the
physical observable at all: $\langle Z_{\rm centre}\rangle$ oscillates by several
tenths about the reference, whereas the $2$-norm curve lies on the
$\chi_{\rm ref}=48$ reference to line width despite its trace defect.  Panel~(c)
quantifies this across bond dimension.  The $2$-norm observable error falls from
$2.3\times10^{-2}$ to $7.6\times10^{-3}$ over $\chi=8\to32$ and is converging,
while the order-adapted error remains of order unity ($1.0\to0.35$), roughly two
orders of magnitude larger at every $\chi$ tested.  The local negativity of the
order-adapted state is non-monotonic in $\chi$ ($0.21,0.42,0.30,0.05$ at
$\chi=8,16,24,32$), suggesting a finite-$\chi$ artifact that vanishes as the
ansatz becomes exact, whereas $2$-norm truncation preserves local positivity
throughout.

The interpretation is straightforward, and it sharpens the point of
Sec.~\ref{app:banach} rather than weakening it.  The base functional selects
components by their pairing with $e$, not by the weight they carry.  Optimizing
that criterion alone therefore discards the dominant Schmidt sectors: it buys an
exactly conserved invariant at the cost of the state itself.  The comparison is
moreover asymmetric in a way that matters practically---the invariant $2$-norm
truncation loses is removable by rescaling, whereas the weight the order-adapted
scheme discards is not.  Preserving the pairing of an ordered dual pair is thus
necessary but not sufficient, and the Banach obstruction is correspondingly
sharper than a change of factorization: a norm-adapted truncation must control
the base norm and the retained weight \emph{together}, which is precisely what
the absent Eckart--Young theorem of Sec.~\ref{app:hilbertian}(c) would supply and
what neither of the two natural criteria supplies alone.

\section{Model Hamiltonians}
\label{app:models}

The first production example is a long-range non-Hermitian Ising chain,
\begin{equation}
  H_{\rm LR\mbox{-}I}
  = J\sum_{i<j}\frac{Z_i Z_j}{|i-j|^\alpha}
    - h\sum_i X_i
    - ik\sum_i Z_i .
  \label{eq:lr_ising}
\end{equation}
For Rydberg arrays in the van der Waals regime one naturally obtains Ising interactions with \(\alpha=6\), while resonant dipolar exchange gives \(\alpha=3\). We use \(\alpha=6\) in the plotted long-range example because it has a direct Rydberg-Ising interpretation and remains numerically tractable as a compressed matrix-product operator~\cite{JavanmardLCU2026}. The small imaginary field \(k\) should be read as a conditional loss/gain term in an effective no-jump Hamiltonian, as discussed in Supplemental Material, Sec.~\ref{app:physical_origin}.

The second production example is the disordered Hatano-Nelson chain,
\begin{align}
  H_{\rm HN} &= t_R \sum_i c^\dagger_{i+1} c_i
               + t_L \sum_i c^\dagger_i c_{i+1}
               + W \sum_i \epsilon_i n_i,
  \label{eq:hn_ferm}
\end{align}
with \(t_{R/L}=t e^{\pm g}\) and quenched disorder \(\epsilon_i\in[-1,1]\). Under the open-chain Jordan-Wigner mapping used in the simulations this becomes
\begin{align}
  H_{\rm HN}
  &= -t\sum_i \left(
      e^g\sigma_i^+\sigma_{i+1}^-
      + e^{-g}\sigma_i^-\sigma_{i+1}^+
    \right)
  + \frac{W}{2}\sum_i \epsilon_i Z_i .
  \label{eq:hn_pauli}
\end{align}
We initialize it in the charge-density-wave state \(|\uparrow\downarrow\uparrow\downarrow\cdots\rangle\) and monitor \(\mathcal{I}(t)=L^{-1}\sum_i(-1)^i\langle n_i(t)\rangle\).

\section{Matrix-Free Local Exponential}
\label{app:matfree}

The local update $\theta'=e^{\dt\,\Heff}\theta$ is evaluated matrix-free by the
repeated scaled Taylor propagation of Algorithm~S1: we estimate $\|\dt\,\Heff\|$
from a few applications of $\Heff$, choose the substep count $s$ so that
$\|\dt\,\Heff\|/s\lesssim 1$, and for each of the $s$ substeps apply the scaled
exponential $e^{\dt\,\Heff/s}$ to the current vector through the truncated Taylor
series of Eq.~\eqref{eq:scaled_taylor}, evaluated entirely with applications of
$\Heff$ via tensor-network contractions. The inner Taylor sum is truncated when
the incremental term falls below a relative tolerance \texttt{tol}, and $s$ is
increased if convergence is not reached within the maximum order $M$. The
effective Hamiltonian $\Heff=L_i^{L(R)}\otimes W_i\otimes W_{i+1}\otimes
R_{i+1}^{L(R)}$ is never assembled as a dense matrix; only its action on a vector
is required. The Taylor tolerance and maximum order used in the runs are
$\texttt{tol}=10^{-12}$ and $M=40$ (Sec.~\ref{app:repro}).

\begin{figure}[t]
\hrule\vspace{2pt}
\textbf{Algorithm S1:} Matrix-free scaled Taylor action
$\theta'\approx e^{\dt\,\Heff}\theta$
\vspace{2pt}\hrule\vspace{3pt}
\begin{algorithmic}[1]
\State estimate $a\leftarrow\|\dt\,\Heff\|$ from a few power applications of $\Heff$
\State $s\leftarrow\max(1,\lceil a\rceil)$;\quad $\theta'\leftarrow\theta$
\For{$j=1$ to $s$}
  \State $\text{term}\leftarrow\theta'$;\quad $\text{acc}\leftarrow\theta'$
  \For{$m=1$ to $M$}
    \State $\text{term}\leftarrow(\dt/s)\,\Heff\,\text{term}/m$ \Comment{one $\Heff$ application}
    \State $\text{acc}\leftarrow\text{acc}+\text{term}$
    \State \textbf{if} $\|\text{term}\|\le\texttt{tol}\,\|\text{acc}\|$ \textbf{then break}
  \EndFor
  \State \textbf{if} not converged at $m=M$ \textbf{then} increase $s$ and restart
  \State $\theta'\leftarrow\text{acc}$
\EndFor
\end{algorithmic}
\vspace{2pt}\hrule
\label{alg:taylor}
\end{figure}

\section{Paired Biorthogonal Two-Site TDVP Sweep}
\label{app:tdvp_alg}

Algorithm~S2 summarizes one time step of the paired left-right two-site TDVP. The
right state is propagated under $H$ and the left state under $H^\dagger$ in
separate two-site sweeps; each local update uses the matrix-free scaled Taylor
action of Algorithm~S1, and the two-site tensor is truncated by an
\emph{independent} SVD at every bond, which additionally grows the bond dimension
analogously to subspace enrichment in DMRG. The single-site back-propagation
between bonds follows the standard two-site TDVP construction~\cite{Haegeman2016},
applied separately to the right and left states; the tensor-network conventions
and pre-trained-MPS initialization follow Ref.~\cite{JavanmardQSM2024}.

\emph{Imaginary-time variant.} Setting $\dt=-\dtau$ in Algorithm~S2 turns the
paired sweep into a double-sided filtering flow. For a diagonalizable,
sufficiently well-conditioned $H$ and an initial pair with nonzero overlap onto
the targeted left and right eigenvectors, the flow amplifies the eigenmode with
the smallest real part of the eigenvalue; the asymptotic approach of the energy
estimate toward $E_0$ is governed by the real gap $\mathrm{Re}(E_1-E_0)$.
Non-normality, eigenvector ill-conditioning, exceptional points, or Jordan
structure can introduce transient growth or polynomial prefactors that modify
this simple exponential picture; none of these regimes is benchmarked in this
work.

\begin{figure}[t]
\hrule\vspace{2pt}
\textbf{Algorithm S2:} Paired biorthogonal two-site TDVP step,
$(\ket{\psi_R},\bra{\psi_L})$ advanced by $\dt$
\vspace{2pt}\hrule\vspace{3pt}
\begin{algorithmic}[1]
\For{$X\in\{R,L\}$ with generator $\{H,\,H^\dagger\}$
  (right under $H$, left under $H^\dagger$)}
  \For{each bond $(i,i{+}1)$ in a left-to-right then right-to-left sweep}
    \State form the two-site tensor $\theta\leftarrow A_X^{s_i} A_X^{s_{i+1}}$
    \State build $L_i^{X},R_{i+1}^{X}$ and
           $\Heff=L_i^{X}\!\otimes W_i\otimes W_{i+1}\otimes R_{i+1}^{X}$
    \State $\theta'\leftarrow e^{\dt\,\Heff}\theta$
           \Comment{Algorithm~S1; $\dt=-i\tau$ (real time)}
    \State SVD $\theta'=U S V^\dagger$; truncate to $\chi_{\max}$;
           renormalize $S$ to unit norm
    \State split into site tensors; back-propagate the bond tensor
           by $e^{-\dt\,\Heff^{(i)}}$
  \EndFor
\EndFor
\State rescale $(\ket{\psi_R},\bra{\psi_L})$ to fix the biorthogonal gauge
\State observables $\langle O\rangle=\langle\psi_L|O|\psi_R\rangle/\langle\psi_L|\psi_R\rangle$
\end{algorithmic}
\vspace{2pt}\hrule
\label{alg:tdvp}
\end{figure}

\section{Hatano-Nelson Spin Representation}
\label{app:hn_decomp}

The compact hopping term used in Eq.~\eqref{eq:hn_pauli},
$-t\sum_i \left(
e^{g}\,\sigma_i^+\sigma_{i+1}^-
+ e^{-g}\,\sigma_i^-\sigma_{i+1}^+
\right)$, can be written explicitly in terms of its real and imaginary parts
using $\sigma^\pm = (\sigma^x \pm i\sigma^y)/2$. Expanding the ladder-operator
products in the text gives
$\sigma_i^+\sigma_{i+1}^- = \tfrac{1}{4}\left[
(\sigma_i^x\sigma_{i+1}^x + \sigma_i^y\sigma_{i+1}^y)
+ i(\sigma_i^y\sigma_{i+1}^x - \sigma_i^x\sigma_{i+1}^y)\right]$
and
$\sigma_i^-\sigma_{i+1}^+ = \tfrac{1}{4}\left[
(\sigma_i^x\sigma_{i+1}^x + \sigma_i^y\sigma_{i+1}^y)
- i(\sigma_i^y\sigma_{i+1}^x - \sigma_i^x\sigma_{i+1}^y)\right]$.
Using $e^{g} + e^{-g} = 2\cosh g$ and $e^{g} - e^{-g} = 2\sinh g$, one finds
\begin{align}
  H_{\rm HN}
  &= -t\sum_i \left(
    e^{g}\,\sigma_i^+\sigma_{i+1}^-
    + e^{-g}\,\sigma_i^-\sigma_{i+1}^+
  \right) \notag \\
  &\qquad = -\frac{t\cosh g}{2}\sum_i
    \bigl(X_i X_{i+1} + Y_i Y_{i+1}\bigr) \notag \\
  &\qquad\quad + \frac{it\sinh g}{2}\sum_i
    \bigl(Y_i X_{i+1} - X_i Y_{i+1}\bigr).
\end{align}
Thus the real part is proportional to $(XX+YY)$ with coefficient
$-t\cosh g/2$, while the imaginary part is proportional to $(YX-XY)$ with
coefficient $it\sinh g/2$.

\section{Overlap-Drift Normalization and Convergence Protocol}
\label{app:health}

\begin{figure}[t]
  \centering
  \includegraphics[width=0.62\columnwidth]{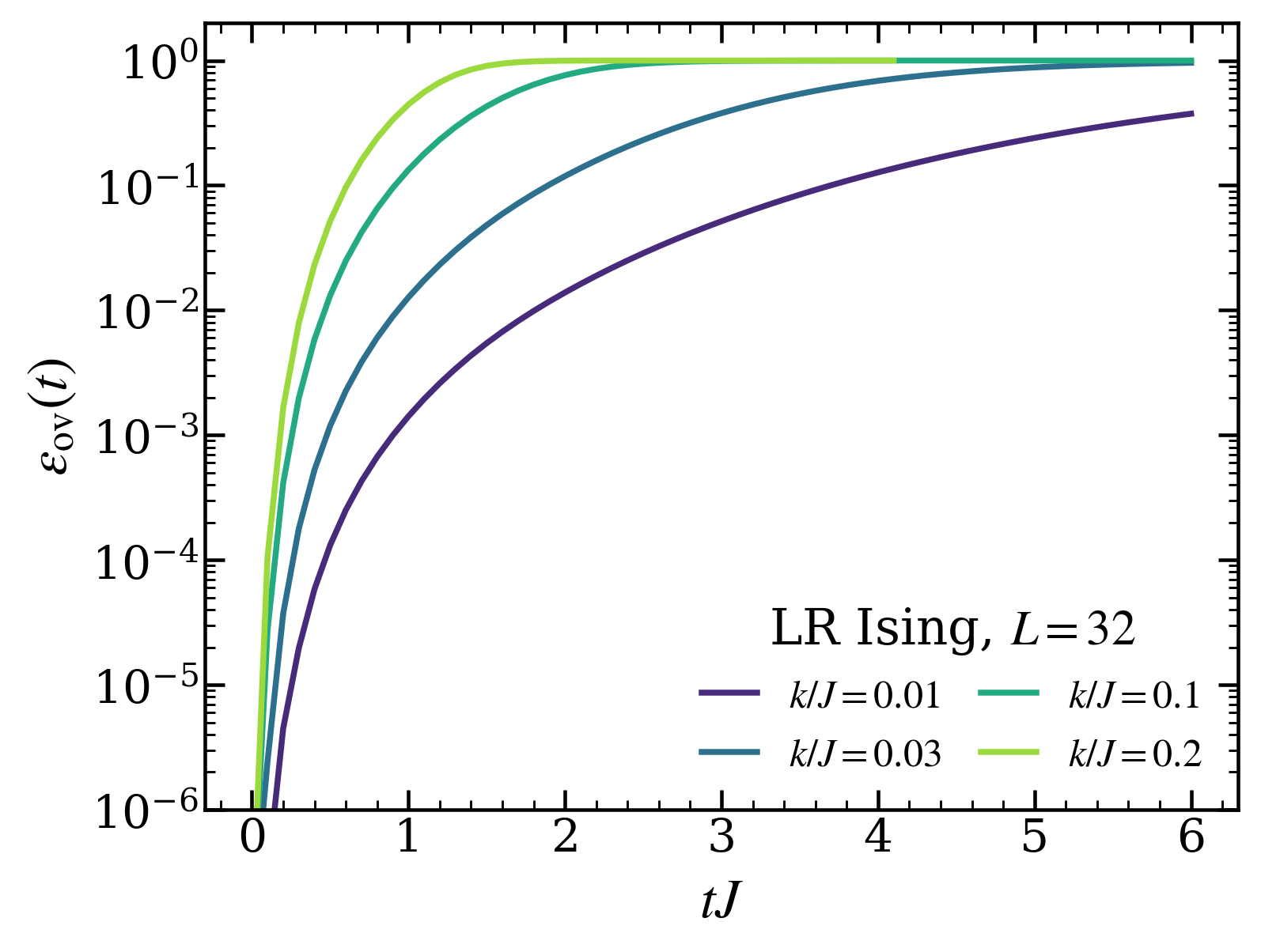}
  \caption{Relative overlap drift of the $L=32$, $\chi=32$ long-range
  non-Hermitian Ising chain for $k/J=0.01,0.03,0.1,0.2$ (dark to light).
  Stronger non-Hermiticity shortens the independently propagated scheme's
  reliable window.}
  \label{fig:kdrift_sm}
\end{figure}

The non-Hermiticity scan in Fig.~\ref{fig:kdrift_sm} shows why the reliable
window must be diagnosed rather than fixed a priori: independent left-right
truncation loses compatibility earlier as the imaginary field is increased.

\emph{Normalization independence of the overlap drift.} To avoid numerical
overflow and underflow, the left and right MPS are normalized during the sweeps:
after every local two-site update and truncation, the Schmidt vector at the
active bond is rescaled to unit norm. Such independent rescalings change the bare
value of $\langle\psi_L|\psi_R\rangle$ even when the propagation and truncation
would preserve it, so the conserved quantity must be reconstructed rather than
read off directly. We therefore retain the cumulative global scale factors
$a_L(t),a_R(t)$ stripped from the two states during normalization and reconstruct
the overlap of the \emph{unnormalized} trajectories,
\begin{equation}
\begin{aligned}
  \mathcal{O}^{\rm raw}_{\rm LR}(t)
  &=a_L(t)\,a_R(t)\,\langle\psi_L(t)|\psi_R(t)\rangle,\\
  \epsilon_{\rm ov}(t)
  &=\left|\frac{\mathcal{O}^{\rm raw}_{\rm LR}(t)}
                 {\mathcal{O}^{\rm raw}_{\rm LR}(0)}-1\right|.
\end{aligned}
  \label{eq:olr_raw}
\end{equation}
which is exactly conserved by the continuous paired dynamics and separates genuine
loss of left--right compatibility from the arbitrary normalization of either MPS.
The drift $\epsilon_{\rm ov}$ of Eq.~\eqref{eq:drift} is computed from this
reconstructed overlap and is therefore independent of the normalization
convention. To isolate the
truncation contribution specifically, one may also record the per-bond drift
induced by each local SVD,
\begin{equation}
  \Delta^{(b)}_{\rm trunc}
  = \frac{\bigl|\mathcal{O}_{\rm LR}^{\rm after}-\mathcal{O}_{\rm LR}^{\rm before}\bigr|}
         {\bigl|\mathcal{O}_{\rm LR}^{\rm before}\bigr|},
  \label{eq:drift_trunc}
\end{equation}
measured immediately before and after truncation at bond $b$, so that the
accumulated $\epsilon_{\rm ov}$ is attributed to independent truncation rather than
to propagation or normalization.

With the biorthonormal
initial product state, $\mathcal{O}^{\rm raw}_{\rm LR}(0)=1$. We do not impose universal
numerical thresholds on $\epsilon_{\rm ov}$: it is an internal-consistency
diagnostic, not an observable-error bound, and its acceptable magnitude is
problem dependent.

The reliable simulation window is instead determined by joint convergence
checks. For a given run we (i) monitor $\epsilon_{\rm ov}(t)$ and the discarded
weight accumulated in the two-site SVDs; (ii) repeat the run at a sequence of
affordable bond dimensions, e.g.\ $\chi=16,24,32$, and require the reported
observables to be stable across the sequence; and (iii) repeat at time steps
$\dt$ and $\dt/2$ and require the observables to be stable under the refinement.
The window is taken to end when any reported observable ceases to be stable under
increasing $\chi$ or decreasing $\dt$. Because $\chi_{\max}=32$ is the largest
bond dimension used here, we state stability within the tested $\chi$ and $\dt$
range rather than full convergence once $\chi$ saturates at $\chi_{\max}$.

\begin{figure}[t]
  \centering
  \includegraphics[width=0.96\columnwidth]{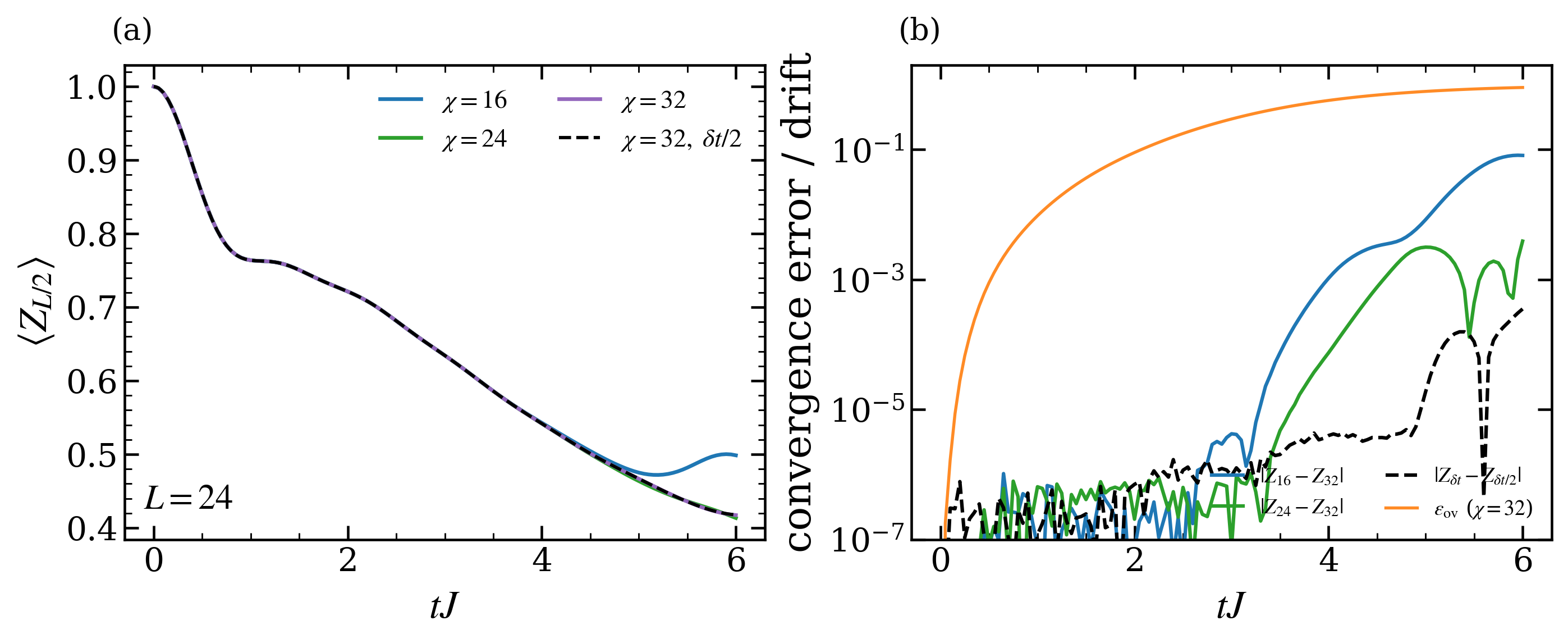}
  \caption{Convergence of the $L=24$ long-range NH Ising production run ($k/J=0.03$). (a) Center-site $\langle Z_{L/2}\rangle(t)$ for bond dimensions $\chi=16,24,32$ at $\dt=0.05$ and for $\chi=32$ at $\dt=0.025$. (b) Convergence errors $|Z_\chi-Z_{32}|$ and the time-step difference $|Z_{\dt}-Z_{\dt/2}|$, together with the relative overlap drift $\epsilon_{\rm ov}(t)$ of the $\chi=32$ run. The reported observable is stable under increasing $\chi$ and decreasing $\dt$ until the convergence errors and $\epsilon_{\rm ov}$ rise together at late times.}
  \label{fig:sm_convergence}
\end{figure}

Figure~\ref{fig:sm_convergence} applies this protocol to the $L=24$ long-range
NH Ising production run at $k/J=0.03$. The center-site observable
[Fig.~\ref{fig:sm_convergence}(a)] is essentially indistinguishable across
$\chi=16,24,32$ and between $\dt$ and $\dt/2$ up to $tJ\approx5$--$6$, and the
pairwise convergence errors [Fig.~\ref{fig:sm_convergence}(b)] stay below
$\sim10^{-3}$ over that interval before rising once $\chi$ saturates and the
trajectories diverge. Notably, the relative overlap drift $\epsilon_{\rm ov}$
grows substantially earlier than the observable convergence errors: it is a
conservative, leading internal-consistency indicator rather than a tight bound on
observable error, consistent with its interpretation in the main text. We
therefore take the reliable window to be the interval over which the observable
is stable under both refinements (here $tJ\lesssim5$--$6$); beyond it, saturation
at $\chi_{\max}=32$ prevents us from claiming convergence.

\paragraph*{Extension to larger chains.}
The production runs use $L=20,24,32$, and the paired left-right TDVP scheme extends
straightforwardly to still larger chains: the per-sweep cost scales roughly
linearly in $L$ at fixed bond dimension, and the matrix-free local propagator is
independent of $L$. The observables and overlap-drift diagnostics are essentially
size-independent across $L=20,24,32$ (Fig.~\ref{fig:tdvp_examples}), so we retain
this range as the production window because it already exposes the accuracy,
convergence, and overlap-drift behavior that are the focus of this work.

\section{Measured Cross-Pairing Conditioning}
\label{app:crosspairing}

The overlap-drift diagnostic $\epsilon_{\rm ov}$ is argued in the main text
(Sec.~\ref{app:banach}) to be a symptom of ill-conditioning of the oblique
projector, i.e.\ of the cross-Gram $G^{LR}$ restricted to the retained bond
subspaces [Eq.~\eqref{eq:banach_infsup}]. Here we measure that object directly, and
then test the attribution by intervention: Sec.~\ref{app:coupled} constrains the
same quantity by construction and shows that the drift falls with it.

\emph{A gauge-invariant local estimate of $\beta$.} At each internal bond $b$
we bring both states to left-canonical form, so that the retained right and left
bond bases $\{|r_i\rangle\}$ and $\{|l_j\rangle\}$ are \emph{orthonormal} on the
left block, and form the $\chi_L\times\chi_R$ overlap matrix
$(M_b)_{ji}=\langle l_j|r_i\rangle$---exactly the retained-subspace block of
$G^{LR}$ that independent truncation perturbs. The canonical form is what makes
this measurement meaningful rather than a basis artifact: it is precisely the
norm-preserving condition required in Sec.~\ref{app:banach} for
$\sigma_{\min}(G^{LR})$ to equal the geometric inf--sup constant, so
$\beta_b=\sigma_{\min}(M_b)$ reports the compatibility of the retained
\emph{subspaces} rather than the conditioning of the tensors used to represent
them. In other words, the construction measures mode (ii) of
Sec.~\ref{app:banach} and is blind by design to the curable parametrization
ill-conditioning of mode (i). The singular values are invariant under the residual
bond gauge, and we report both $\beta_b$ and the conditioning
$\kappa(M_b)=\sigma_{\max}(M_b)/\beta_b$.  Both states use the same target
$\chi_{\max}$; the square cross matrices reported below are evaluated on bonds
with matched retained dimensions.  This matters beyond convenience: by the
remark following Eq.~\eqref{eq:banach_oblique_projector}, a rectangular block
with $\chi_R>\chi_L$ would give $\beta_b=0$ identically, and the measurement
would report degeneracy that is an artifact of unequal retained dimensions rather
than of lost compatibility.

\begin{figure}[t]
  \centering
  \includegraphics[width=\columnwidth]{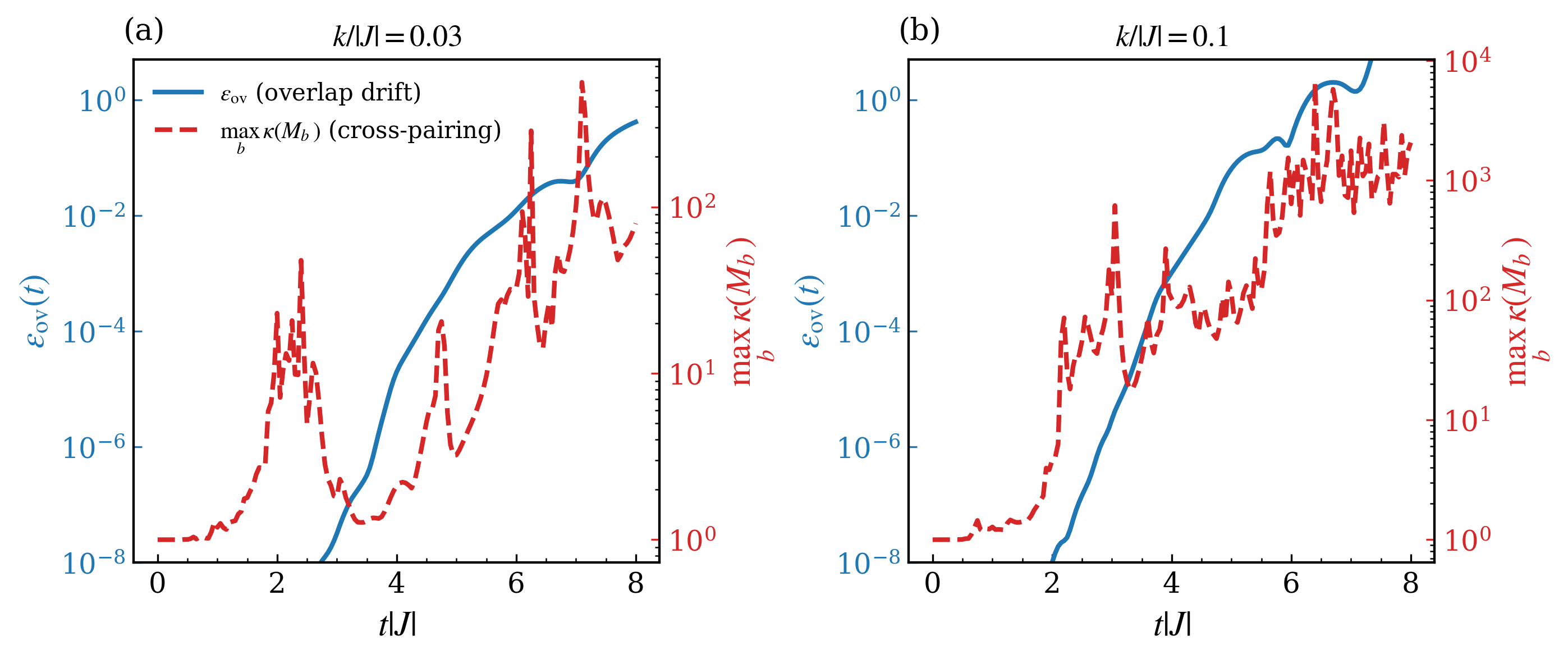}
  \caption{Measured cross-pairing conditioning versus overlap drift for the
  $L=20$ long-range NH Ising run ($J=-1$, $h=0.65$, $\alpha=6$, $\chi=32$).
  Overlap drift $\epsilon_{\rm ov}(t)$ (solid, left axis) and the worst-bond
  conditioning $\max_b\kappa(M_b)$ of the retained cross-Gram (dashed, right axis)
  for (a) $k/|J|=0.03$ and (b) $k/|J|=0.1$. The two rise together, and the
  associated inf--sup constant $\beta_b=\sigma_{\min}(M_b)$ falls to $\sim10^{-5}$
  where the drift sets in. Because both quantities grow monotonically over the
  window, the causal attribution rests on the intervention of
  Fig.~\ref{fig:coupled_main} rather than on this co-occurrence: constraining
  $\kappa(M_b)$ lowers $\epsilon_{\rm ov}$.}
  \label{fig:crosspairing}
\end{figure}

\emph{Measured behaviour.} Figure~\ref{fig:crosspairing} tracks
$\max_b\kappa(M_b)$ alongside $\epsilon_{\rm ov}(t)$ for the $L=20$ long-range NH
Ising run. The two rise together: as $\epsilon_{\rm ov}$ grows from $10^{-3}$
toward order unity, the worst-bond conditioning grows by two to three orders of
magnitude and the inf--sup constant collapses to $\beta_b\sim10^{-5}$, so the
retained cross-pairing becomes nearly singular over the same interval in which the
overlap drifts.

\emph{The attribution is established by intervention, not by co-occurrence.}
That the two curves rise together is suggestive but not conclusive, since both
grow monotonically over the window. The decisive test is to constrain
$\kappa(M_b)$ and ask what happens to the drift. This is exactly what the coupled
biorthonormal truncation of Sec.~\ref{app:coupled} does: run on the same
trajectory with the same bond dimension, it holds $\max_b\kappa(M_b)$ near
$\mathcal{O}(1$--$10)$ where the independent scheme spikes to $10^{3}$, and the
overlap drift falls by about an order of magnitude in consequence
(Fig.~\ref{fig:coupled_main}). Suppressing the cross-pairing degeneracy therefore
suppresses the drift, which identifies the degeneracy of $G^{LR}$---rather than
norm growth or basis conditioning---as what $\epsilon_{\rm ov}$ detects.

\section{Coupled Biorthogonal Truncation}
\label{app:coupled}

Section~\ref{app:crosspairing} identifies the degeneracy of the retained
cross-Gram as the origin of the overlap drift. This motivates a truncation that
keeps the two bond bases in duality by construction. We implement and test it
here. At each bond we form the (non-Hermitian) cross density matrix
\begin{equation}
  \rho_b=\mathrm{Tr}_{\rm env}\,|\psi_R\rangle\langle\psi_L|
        = W_R\,\mathcal{E}^{\!\top}\,W_L^{\dagger},
  \label{eq:cross_dm}
\end{equation}
where $W_R, W_L$ are the left parts of the two states at the bond and $\mathcal{E}$
is the right cross-environment; this is the time-dependent analogue of the
biorthonormal-block reduced density matrix of Ref.~\onlinecite{Zhong2025}. Its
right and left eigenvectors furnish a biorthonormal pair of bond bases
($\langle l_j|r_i\rangle=\delta_{ij}$, so $M_b=\mathbb{1}$ by construction); we
keep the $\chi_{\max}$ eigenvalues of largest magnitude, and fall back to
independent singular value truncation whenever the eigenvector matrix is too
ill-conditioned to trust ($\kappa>\kappa_{\max}$), the conditioning control of
Ref.~\onlinecite{Zhong2025}. The construction is exact at full bond dimension
(it reproduces the cross-overlap to machine precision) and, under truncation,
preserves the cross-overlap markedly better than independent truncation; a
random-pair sanity check at $\chi=6$ gives an overlap error smaller by more than
an order of magnitude.

Figure~\ref{fig:coupled_main} runs the two schemes dynamically on the $L=20$
long-range NH Ising trajectory at $k/|J|=0.1$ and $\chi=16$. To isolate the
truncation operator from every other difference between the schemes, both are
applied to the \emph{same} trajectory: the state is evolved at a larger reference
bond dimension $\chi_{\rm ref}=32$ and compressed to $\chi$ after each step, so
the comparison is between two compressions of an identical evolution rather than
between two independent runs. The coupled
truncation holds the overlap drift $\epsilon_{\rm ov}$ about an order of magnitude
below independent truncation across the reliable window, and keeps the worst-bond
conditioning $\max_b\kappa(M_b)$ near $\mathcal{O}(1$--$10)$ where the independent
scheme repeatedly spikes to $10^{3}$. Both eventually grow once the paired state
becomes genuinely hard at late times, but the coupled scheme extends the window
over which the left/right pairing stays well conditioned. This is the intervention
that underwrites the attribution of Sec.~\ref{app:crosspairing}: constraining
$\kappa$ by construction lowers the measured drift. The coupled truncation is
thus a demonstrated remedy, at the cost of the non-Hermitian bond eigendecomposition
and its conditioning control.

\section{Coupled Oblique Biorthogonal TDVP}
\label{app:coupled_update}
\suppressfloats[t]

Sections~\ref{app:crosspairing}--\ref{app:coupled} address the truncation. The
local \emph{update} can also be made genuinely coupled, closing the gap between the
oblique projection of Sec.~\ref{sec:method} and the paired two-site algorithm. In the
coupled two-site update the right tensor evolves by
\begin{equation}
  i\,\dot\theta_R = G_L^{-1}\, H^{\rm cross}_{\rm eff}[\psi_L,\psi_R]\,\theta_R\, G_R^{-1},
  \label{eq:coupled_update}
\end{equation}
where the cross effective Hamiltonian $H^{\rm cross}_{\rm eff}$ is built with the
\emph{left} state as bra and the \emph{right} state as ket---so the right update is
tested against the left tangent space---and $G_L,G_R$ are the left/right cross-Gram
blocks $M_b$ of Sec.~\ref{app:crosspairing}. The explicit inversions $G_L^{-1},G_R^{-1}$
are the cross-pairing (oblique) solve; the left tensor evolves symmetrically under
$H^\dagger$ with bra and ket exchanged. Equation~\eqref{eq:coupled_update} is the direct
discretization of the projected equations of motion,
Eqs.~\eqref{eq:tdvp_eom}--\eqref{eq:metric}: the partner state enters the local
effective Hamiltonian, and the cross-pairing is solved rather than assumed diagonal.

\begin{figure}[!b]
  \centering
  \includegraphics[width=\columnwidth]{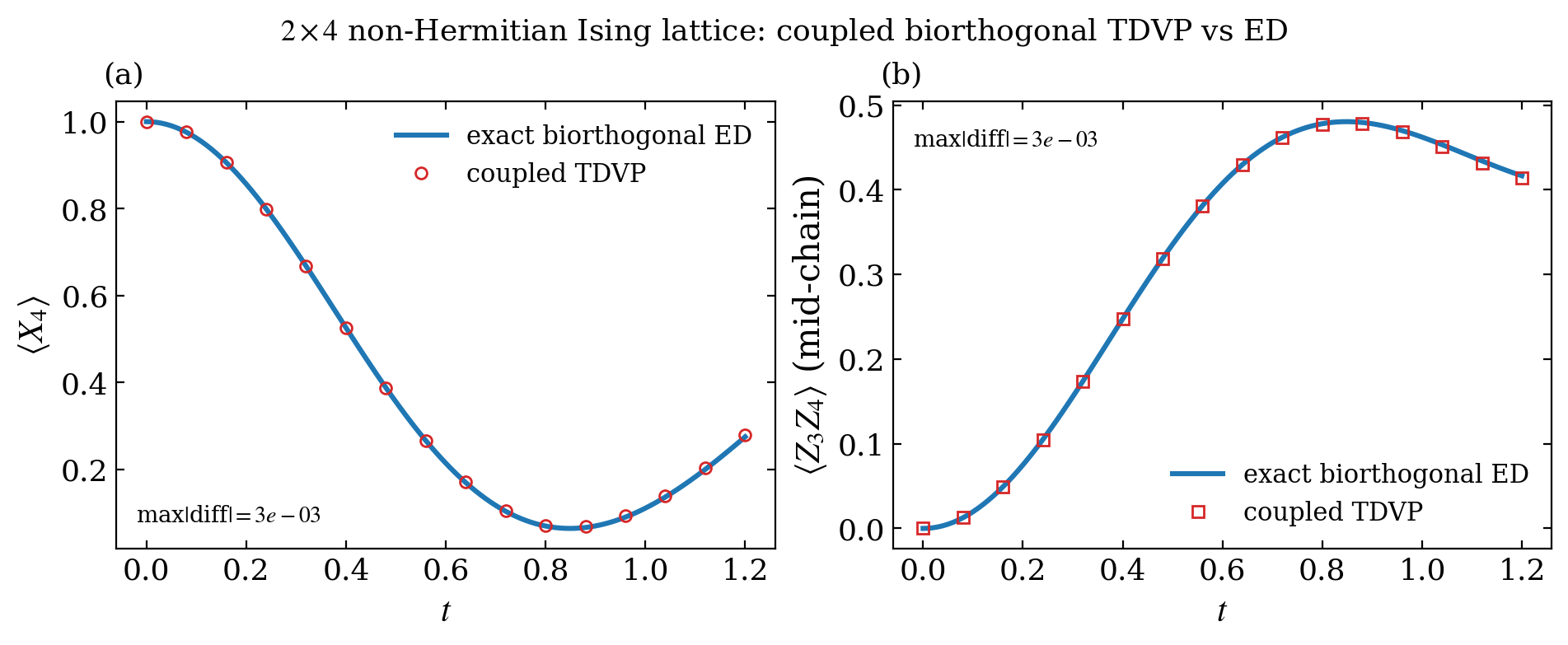}
  \caption{Coupled oblique biorthogonal TDVP versus exact biorthogonal
  diagonalization for a $2\times4$ non-Hermitian Ising lattice (snake-mapped MPS).
  (a) Center-site $\langle X\rangle$ and (b) mid-chain correlation
  $\langle Z_aZ_b\rangle$: the coupled TDVP (markers) tracks exact ED (lines) to the
  time-step error. The local update uses the cross effective Hamiltonian and the explicit
  $G^{LR}$ solve of Eq.~\eqref{eq:coupled_update}.}
  \label{fig:coupled_2d}
\end{figure}

We implement this coupled update as a symmetric (second-order) two-site integrator
and validate it against exact biorthogonal diagonalization. For the one-dimensional
non-Hermitian transverse-field Ising chain it reproduces the exact $\langle X_{L/2}\rangle$
and the two-point functions $\langle Z_aZ_b\rangle$ to $10^{-12}$ at full bond dimension;
At full bond dimension, the tangent space is the whole space, so this validates the
integrator, the cross environments, and the matrix-free exponential rather than the
oblique projection itself. The
scheme is geometry agnostic: a two-dimensional lattice maps to a one-dimensional
matrix-product state along a snake path, the vertical bonds becoming finite-range
couplings that only widen the matrix-product operator (whose generic construction we
check reproduces the dense Hamiltonian exactly). Figure~\ref{fig:coupled_2d} shows a
$2\times4$ non-Hermitian Ising lattice: the coupled TDVP reproduces the exact center-site
$\langle X\rangle$ and the mid-chain correlation $\langle Z_aZ_b\rangle$ to within the
time-step error. The coupled oblique dynamics of Sec.~\ref{sec:method} is therefore not
only derived but implemented and correct; the decoupled scheme used for the large-scale
production runs is its efficient approximation, whose departure is quantified in
Sec.~\ref{app:crosspairing} and remedied at the truncation level in
Sec.~\ref{app:coupled}.

\paragraph*{Cross environments.} The coupled update needs, at each bond, the
effective Hamiltonian and the cross-Grams built from \emph{both} states. Writing
$A^{[i]}$ for the right-state site tensor (ket) and $B^{[i]}$ for the left-state site
tensor (bra), the left cross environment carrying the MPO tensor $W^{[i]}$ satisfies the
transfer recursion
\begin{equation}
  (L_{i+1})^{w'}_{a'_L a'_R}
  = \sum \bigl(B^{[i]}\bigr)^{*\,p}_{a_L a'_L}\,(L_i)^{w}_{a_L a_R}\,
        \bigl(W^{[i]}\bigr)^{w w'}_{p\,q}\,\bigl(A^{[i]}\bigr)^{q}_{a_R a'_R},
  \label{eq:cross_env}
\end{equation}
and the right cross environment $R_i$ obeys the mirror recursion from the right boundary.
The cross-Gram blocks are the same contractions with the MPO removed,
$(G_{L,i})_{a'_L a'_R}=\sum (B^{[i]})^{*p}_{a_L a'_L}(G_{L,i-1})_{a_L a_R}(A^{[i]})^{p}_{a_R a'_R}$,
so that $G_{L,i}=M_i$ is precisely the retained cross-Gram of Sec.~\ref{app:crosspairing}.
The two-site cross effective Hamiltonian in Eq.~\eqref{eq:coupled_update} is the
contraction $L_i$--$W^{[i]}$--$W^{[i+1]}$--$R_{i+2}$ applied to $\theta_R$; it is a
non-normal operator, applied matrix-free.

\paragraph*{Coupled two-site integrator.} One time step is a symmetric
(second-order) two-site sweep. Left to right, at each bond $(i,i+1)$ we (i) build the
cross environments and cross-Grams from the current pair, (ii) evolve the right two-site
tensor by $\theta_R\!\leftarrow\!\exp(-\tfrac{i}{2}\,\delta t\,G_L^{-1}H^{\rm cross}_{\rm eff}G_R^{-1})\,\theta_R$
and the left two-site tensor symmetrically under $H^\dagger$ with bra and ket exchanged,
(iii) split each state by its own singular value decomposition, advancing the
orthogonality center, and (iv) back-evolve the resulting one-site center by
$-\tfrac{1}{2}\delta t$ with the corresponding one-site coupled generator. The reverse
sweep completes the step. The matrix exponentials are applied by the same matrix-free
scaled Taylor action used in the main text (Sec.~\ref{app:matfree}), so the coupled
generator $G_L^{-1}H^{\rm cross}_{\rm eff}G_R^{-1}$ is only ever \emph{applied}, never
formed. Note that the substep count of Algorithm~S1 is set from a norm estimate of that
coupled generator, which carries a factor $\|G^{-1}\|\sim1/\beta_b$: the inf--sup
constant of Sec.~\ref{app:crosspairing} therefore enters the runtime directly, and the
biorthonormal gauge of Sec.~\ref{app:coupled}, in which $G_L=G_R=\mathbb{1}$, removes
that factor as well as the inversion. The cross-Grams are otherwise inverted directly;
where $\beta_b$ is small a Tikhonov ridge $G\!\to\!G+\eta\mathbb{1}$ regularizes the
solve.

\paragraph*{Two dimensions.} A two-dimensional $L_x\times L_y$ lattice is mapped to
a chain by a boustrophedon (snake) path; nearest-neighbor vertical bonds become
finite-range couplings of range $\sim L_y$, and the coupled update is otherwise
unchanged. We build the required matrix-product operator generically from the list of
onsite and two-site terms by a finite-state construction: each two-site term
$J\,O_a O_b$ opens a channel at site $a$ (transition from the start state with weight
$J O_a$), carries the identity on the intervening sites, and closes at site $b$
(transition to the accumulated state with $O_b$); a greedy interval coloring reuses
channels so that the operator bond dimension equals two plus the maximal number of bonds
crossing a cut, i.e.\ $\sim L_y$. The construction is verified to reproduce the dense
Hamiltonian exactly (to machine precision) for every lattice tested. These lattices are
small enough to be reachable by exact diagonalization; they demonstrate that the coupled
update is correct in two dimensions, not that it reaches sizes exact methods cannot.

\begin{figure}[t]
  \centering
  \includegraphics[width=0.8\columnwidth]{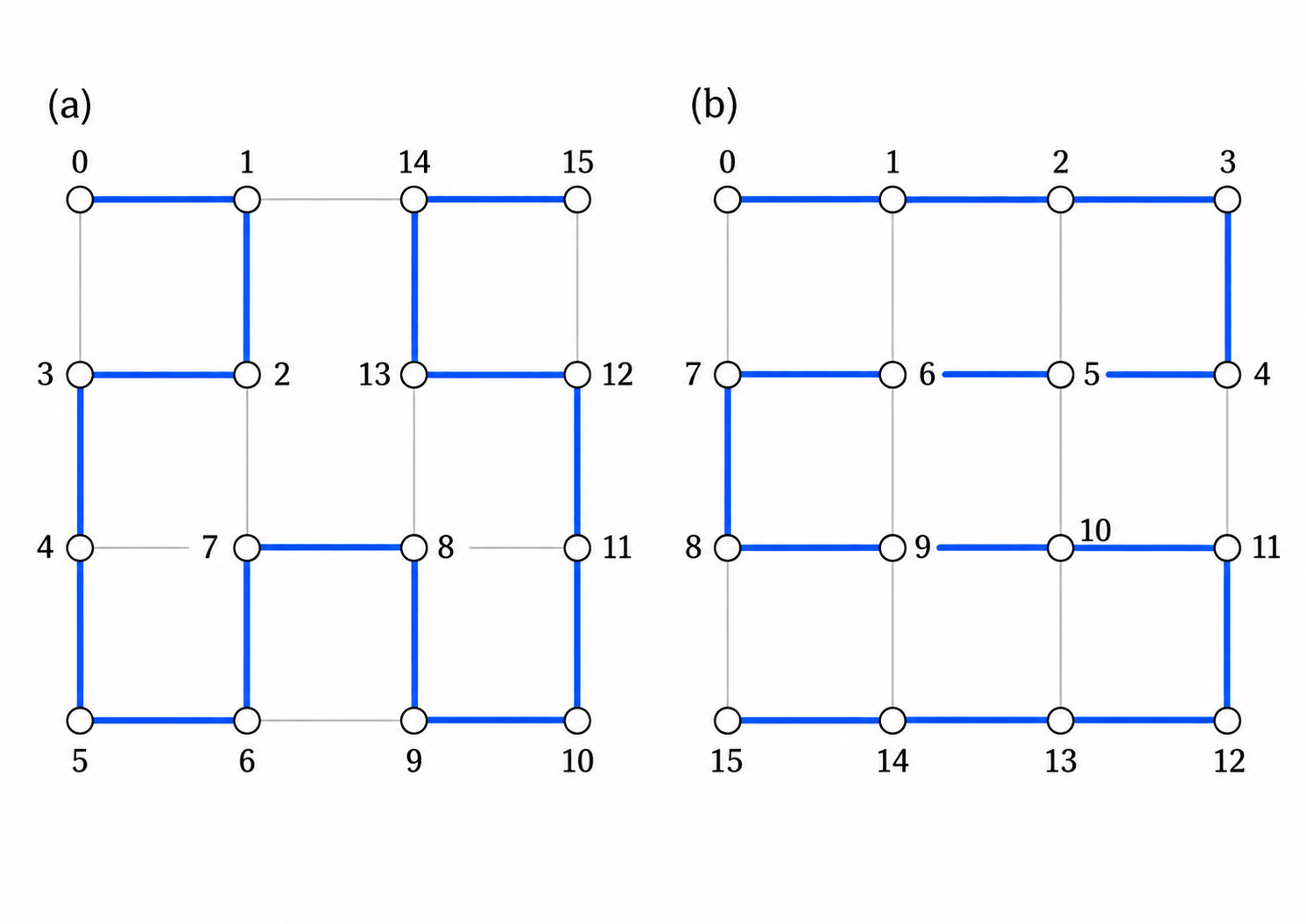}
  \caption{Qubit linearization of the $4\times4$ lattice for the coupled
  biorthogonal TDVP. A Hilbert space-filling curve (a) keeps each contiguous segment of
  the matrix-product state inside a compact two-dimensional region, lowering the
  entanglement across cuts relative to the boustrophedon (snake) path (b). Squares mark
  the start and diamonds the end of each path.}
  \label{fig:hilbert}
\end{figure}

\begin{figure}[t]
  \centering
  \includegraphics[width=\columnwidth]{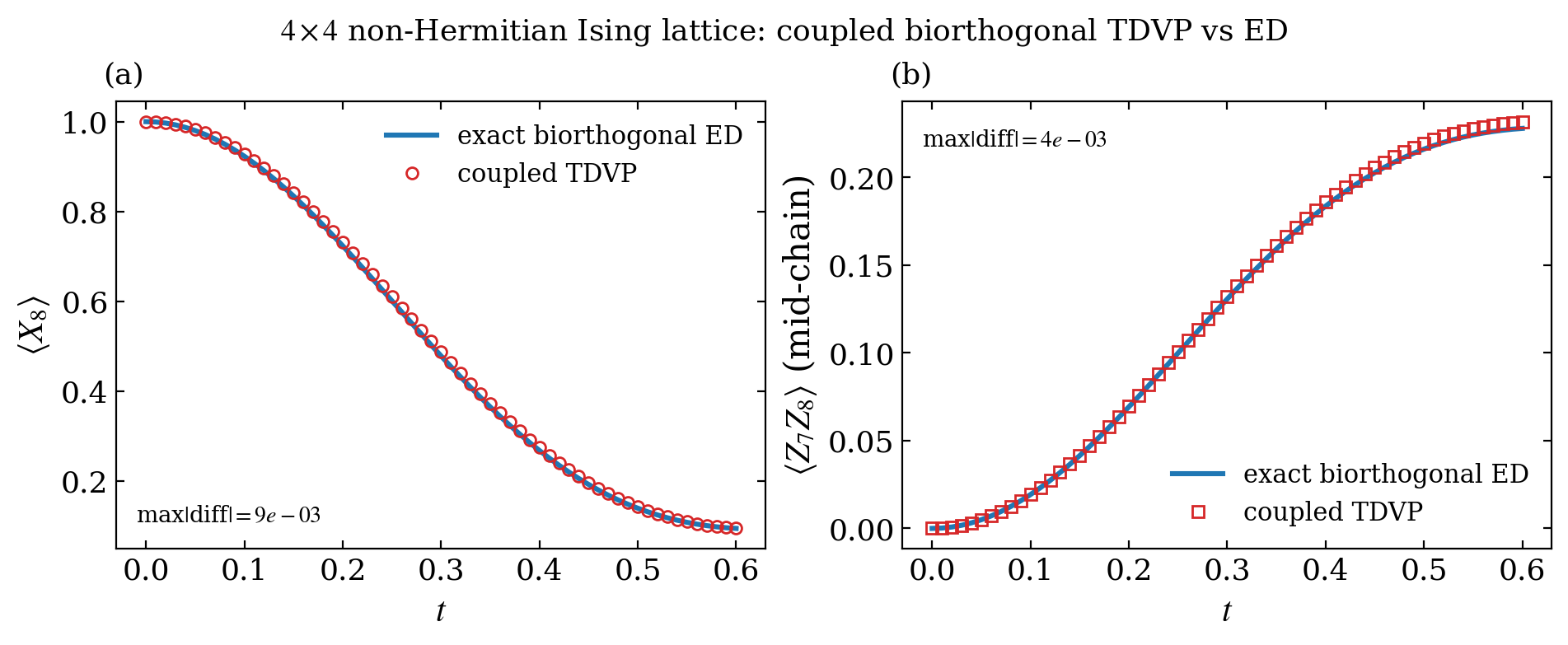}
  \caption{Coupled oblique biorthogonal TDVP on a $4\times4$ ($16$-site)
  non-Hermitian Ising lattice ($k=0.08$, Hilbert-curve linearization), versus exact
  biorthogonal diagonalization by a sparse Krylov propagator. (a) Center-site
  $\langle X\rangle$ and (b) mid-chain $\langle Z_aZ_b\rangle$: the coupled TDVP (markers)
  reproduces exact ED (lines) to the time-step error ($9\times10^{-3}$ and
  $4\times10^{-3}$ at $\dt=0.01$). The run uses the jointly biorthonormal bond basis and
  the fault-tolerant SVD described in the text.}
  \label{fig:coupled_4x4}
\end{figure}

For the square $4\times4$ lattice we instead linearize the qubits along a
\emph{Hilbert space-filling curve} (Fig.~\ref{fig:hilbert}): a contiguous segment of the
chain then maps to a compact two-dimensional region, which lowers the entanglement across
the matrix-product cuts relative to the snake path and keeps the required bond dimension
under control. Two further ingredients make the coupled update robust in this regime.
First, the two-site tensors are split with a fault-tolerant singular value decomposition:
if LAPACK's divide-and-conquer driver fails to converge---which can happen when the
cross-Gram update produces a large dynamic range near a biorthogonal singularity---we fall
back to a plain decomposition and finally to the eigen-decomposition of $M^\dagger M$,
which always converges. Second, rather than splitting each state independently and
inverting the cross-Grams $G_L,G_R$ at every bond, we split the pair in a \emph{jointly
biorthonormal} bond basis (Sec.~\ref{app:coupled}), in which $G_L=G_R=\mathbb{1}$ and no
inversion is needed; a conditioning fallback reverts to independent splits where the
biorthonormal basis is ill-defined. This removes the spurious bond-dimension growth that
an ill-conditioned $G^{-1}$ would otherwise inject and lets the $4\times4$ run proceed
stably to the truncation-limited accuracy of Fig.~\ref{fig:coupled_4x4}.

\paragraph*{Exact-diagonalization reference.} The exact biorthogonal reference
propagates $|\psi_R\rangle$ under $e^{-iHt}$ and $|\psi_L\rangle$ under $e^{-iH^\dagger t}$
and evaluates $\langle\psi_L|O|\psi_R\rangle/\langle\psi_L|\psi_R\rangle$. For $L\lesssim12$
the propagator is a dense matrix exponential; for the $4\times4$ lattice ($L=16$, a
$2^{16}$-dimensional space for which a dense Hamiltonian is infeasible) it is a sparse
Krylov exponential action, $e^{-iH\,\delta t}$ applied by \texttt{expm\_multiply} to a
sparse $H$. Across one- and two-dimensional lattices the coupled TDVP reproduces this
reference to the time-step error, confirming that the oblique update of
Eq.~\eqref{eq:coupled_update} implements the projected dynamics of Sec.~\ref{sec:method}
exactly at full bond dimension.

\paragraph*{Coupled update for the production DQPT.} The dynamical quantum phase
transitions of Sec.~\ref{sec:dqpt} are the most demanding production observable. It is
worth being precise about why. The Fisher zero is a zero of the \emph{return} amplitude
$\langle\psi_L(0)|\psi_R(t)\rangle$, not of the equal-time pairing
$\langle\psi_L(t)|\psi_R(t)\rangle$, which the paired flow conserves; a DQPT does not
therefore by itself render the cross-Gram singular. What it does render ill-conditioned is
the estimator: $\lambda_{\rm bi}$ is the logarithm of an amplitude that passes through
zero, so its relative error near the cusp is amplified by $1/|F(t)|$, and both factors of
$F$ are exponentially small in $L$ and obtained by cancellation.  The
implementation therefore provides a Tikhonov ridge $G\to G+\eta\mathbb{1}$ as a
conditioning fallback when the measured per-bond inf--sup constant $\beta_b$ of
Sec.~\ref{app:crosspairing} becomes small.
Figure~\ref{fig:dqpt_coupled_main} shows that the coupled biorthogonal rate reproduces exact
biorthogonal diagonalization \emph{through} the cusps, to $2$--$4\times10^{-5}$ at
$k=0.05$ and $0.15$ ($L=10$), and tracks the biorthogonal rate where it departs sharply
from the conditional (single-branch) one. The reported DQPT is therefore not an artifact of
a decoupled truncation: the genuinely coupled oblique update---partner state in the local
test space, cross-Gram solved rather than assumed diagonal---yields the same transition and
the same exact rate.

\begin{figure}[t]
  \centering
  \includegraphics[width=\columnwidth]{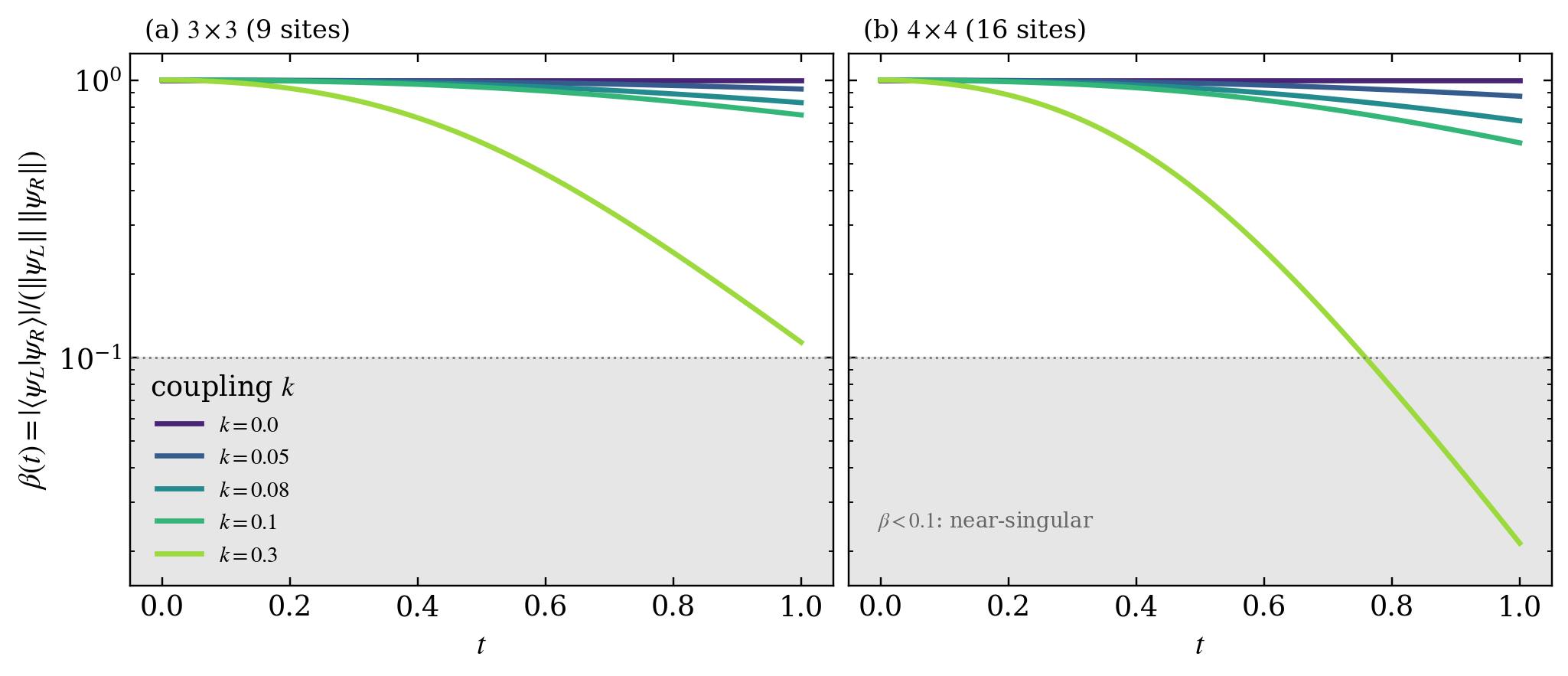}
  \caption{Biorthogonal overlap $\beta(t)$ of Eq.~\eqref{eq:beta_global_main} versus the
  non-Hermitian coupling $k$, for (a) a $3\times3$ ($9$-site, dense propagator) and (b) a
  $4\times4$ ($16$-site, sparse Krylov propagator) non-Hermitian Ising lattice. At $k=0$ the
  dynamics is unitary and $\beta\equiv1$; increasing $k$ (and system size) drives $\beta$
  down exponentially. The shaded band $\beta<0.1$ marks the regime in which the conserved
  pairing must be reconstructed from states whose norms multiply to $\gtrsim10$, so that
  relative errors in either state are amplified by $1/\beta$ in every biorthogonal
  expectation value. The production coupling
  $k<0.1$ keeps $\beta=\mathcal{O}(1)$ over the simulated window, the regime in which the
  coupled oblique TDVP of Eq.~\eqref{eq:coupled_update} reproduces exact biorthogonal
  diagonalization.}
  \label{fig:beta_coupling}
\end{figure}

\paragraph*{Biorthogonal overlap and the role of the coupling.} The tractability of
the coupled dynamics is governed by the global inf--sup constant $\beta(t)$ of
Eq.~\eqref{eq:beta_global_main}, the rank-one specialization of
Eq.~\eqref{eq:banach_infsup} identified in the main text, of which the per-bond
$\beta_b=\sigma_{\min}(M_b)$ of Sec.~\ref{app:crosspairing} is the
retained-subspace restriction.
Because the \emph{unnormalized} biorthogonal overlap is
conserved under the paired flow, $\langle\psi_L(t)|\psi_R(t)\rangle=
\langle\psi_0|e^{iHt}e^{-iHt}|\psi_0\rangle=\mathrm{const}$, the decay of $\beta(t)$ is
entirely due to the growth of the individual norms,
$\beta(t)=1/(\|\psi_L\|\,\|\psi_R\|)$, which is a purely non-normal effect: for a
Hermitian generator ($k=0$) the evolution is unitary and $\beta\equiv1$ for all time.
Switching on the non-Hermitian coupling $k$ gives the eigenvalues imaginary parts
$\sim k$; different modes are amplified at different rates, the norms grow, and $\beta(t)$
falls---exponentially in $k$ and in the system size, as shown in
Fig.~\ref{fig:beta_coupling}.

The coupled TDVP reproduces exact biorthogonal ED wherever $\beta$ is
$\mathcal{O}(1)$ (to $10^{-12}$ in one dimension, to the time-step error in two).
Small $\beta$ does \emph{not} make the biorthogonal estimator singular---its denominator
is conserved---but it does make it ill-conditioned. When $\beta\ll1$ an
$\mathcal{O}(1)$ pairing is reconstructed from two states whose norms multiply to
$1/\beta$, so a relative perturbation $\varepsilon$ of either state perturbs
$\langle\psi_L|O|\psi_R\rangle/\langle\psi_L|\psi_R\rangle$ by
$\mathcal{O}(\varepsilon/\beta)$. This is the same amplification factor that bounds the
oblique projector in Eq.~\eqref{eq:proj_bound}, here at the level of the global pair
rather than of the tangent spaces.
The production runs therefore sit at moderate
coupling $k<0.1$, which keeps $\beta=\mathcal{O}(1)$ over the simulated window even at the
largest sizes (Fig.~\ref{fig:beta_coupling}b): $\beta(t\!=\!1)\!\approx\!0.71$ at $k=0.08$
for the $4\times4$ lattice, versus $0.02$ at $k=0.3$. Equation~\eqref{eq:beta_global_main} thus
serves as an \emph{a priori} diagnostic that predicts, from the coupling and the elapsed
time alone, where the coupled biorthogonal scheme is well conditioned.

\FloatBarrier
\section{Additional Benchmark Models}
\label{app:additional_models}

The main text focuses on the long-range non-Hermitian Ising chain and the disordered Hatano-Nelson chain. Several additional models were used to validate the implementation against exact diagonalization (ED).

\begin{figure}[t]
  \centering
  \includegraphics[width=0.96\columnwidth]{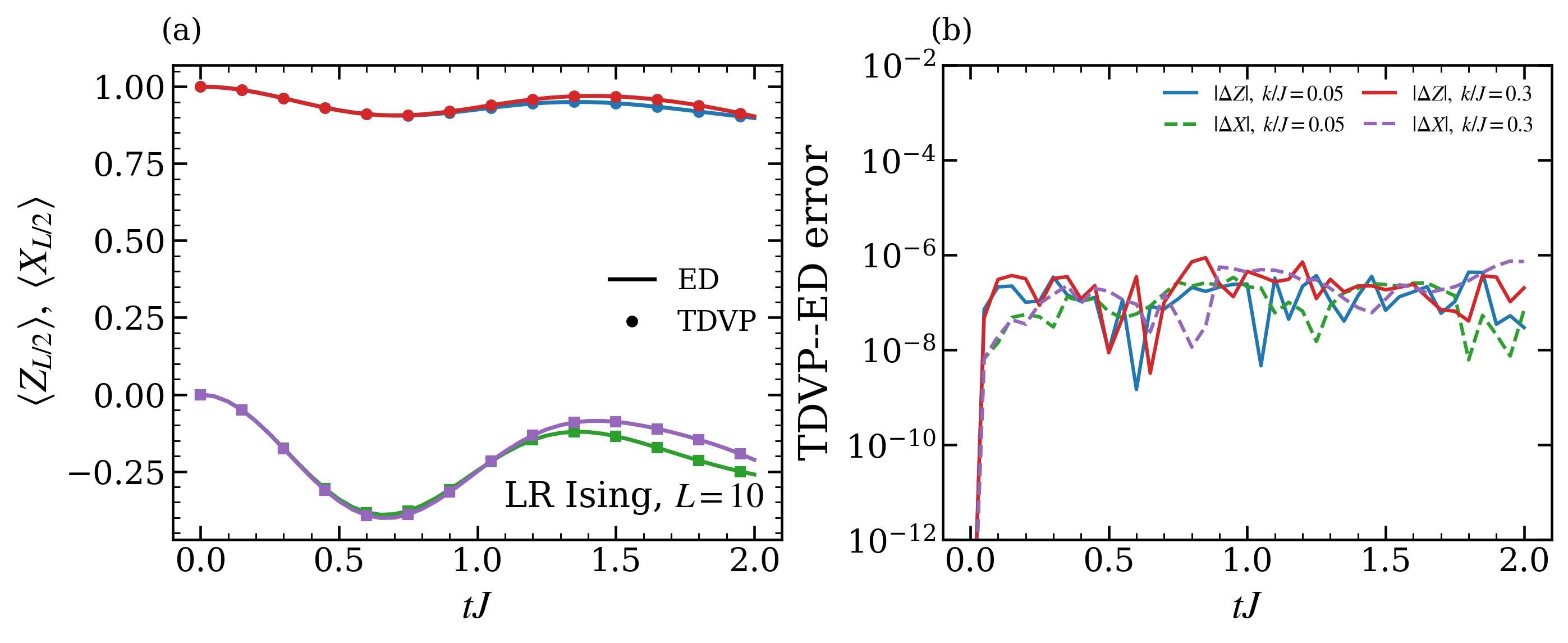}
  \caption{Exact diagonalization versus biorthogonal TDVP for the long-range NH Ising chain ($L=10$, $\alpha=3$). (a) Center-site $\langle Z_{L/2}\rangle$ and $\langle X_{L/2}\rangle$: exact (lines) and TDVP (markers) for weak ($k/|J|=0.05$) and strong ($k/J=0.3$) non-Hermitian fields. (b) Absolute TDVP--ED error of the same observables.}
  \label{fig:sm_ed_vs_tdvp}
\end{figure}

Figure~\ref{fig:sm_ed_vs_tdvp} compares the biorthogonal TDVP trajectories directly with exact biorthogonal evolution for the long-range NH Ising chain at $L=10$, where exact evolution is feasible. The center-site $\langle Z_{L/2}\rangle$ and $\langle X_{L/2}\rangle$ from TDVP are indistinguishable from ED for both weak ($k/|J|=0.05$) and strong ($k/J=0.3$) non-Hermitian fields, with absolute errors that remain small over the evolution.

\paragraph*{Nearest-neighbor NH-TFIM.}
The short-range non-Hermitian transverse-field Ising benchmark is
\begin{equation}
  H_{\rm TFIM}=-\lambda\sum_i Z_iZ_{i+1}-h\sum_i X_i-ik\sum_i Z_i .
\end{equation}
It is useful for direct small-system validation because ED is cheap up to \(L\simeq 10\): two-site MPS TDVP agrees with ED at \(L=6,8\) to below \(10^{-11}\) for center-site observables.

\paragraph*{Long-range dipolar Ising and XY chains.}
For dipolar exchange platforms one may use \(\alpha=3\) rather than the van der Waals value \(\alpha=6\):
\begin{align}
  H_{\rm LR\mbox{-}I}^{(3)}
  &= J\sum_{i<j}\frac{Z_iZ_j}{|i-j|^3}-h\sum_i X_i-ik\sum_i Z_i,\\
  H_{\rm LR\mbox{-}XY}^{(3)}
  &= \frac{J}{2}\sum_{i<j}\frac{X_iX_j+Y_iY_j}{|i-j|^3}-ik\sum_i Z_i .
\end{align}
For the Ising version, \(L=8,10\) ED comparisons have maximum \(Z_{L/2}\) and \(X_{L/2}\) errors below \(10^{-5}\). The XY form probes exchange-dominated transport rather than ordering dynamics.

\paragraph*{Lindblad amplitude damping.}
The stochastic open-system benchmark uses jump operators \(L_i=\sqrt{\gamma}\,\sigma_i^-\) and the Lindblad equation in Eq.~\eqref{eq:lindblad}. Its no-jump generator, Eq.~\eqref{eq:amp_damp_heff}, is evolved by the same non-Hermitian TDVP kernel. Averaging quantum-jump trajectories reconstructs full Lindblad expectation values, while the deterministic no-jump branch tests TDVP accuracy for \(H_{\rm eff}\).

\section{Biorthogonal DQPT Construction on an Exactly-Solvable Model}
\label{app:ssh_dqpt}

The main-text DQPT analysis (Sec.~\ref{sec:dqpt}) uses the biorthogonal rate
$\lambda_{\rm bi}$ of Eq.~\eqref{eq:rate_function}, built from the paired left/right
states. To cross-check that biorthogonal construction against an independent route
---the momentum-space associated-state inner product on which it rests---we apply
it to a model where a momentum-resolved analysis is exact, the non-Hermitian
Su--Schrieffer--Heeger (SSH) chain of Ref.~\onlinecite{BiorthDQPT2024}. Its Bloch
Hamiltonian is $H_k=\mathbf{d}_k\!\cdot\!\boldsymbol{\sigma}$ with
$\mathbf{d}_k=\bigl((1+\eta)+(1-\eta)\cos k,\ (1-\eta)\sin k-\mathrm{i}\gamma/2,\ 0\bigr)$,
where $\eta$ tunes the intra/inter-cell hopping and $\gamma$ the non-Hermiticity;
the dispersion $\pm\varepsilon_k=\pm\sqrt{d_x^2+d_y^2}$ is complex.

For each momentum we build the associated state of
Ref.~\onlinecite{nonHer004_brody2013biorthogonal}
in the post-quench right eigenbasis. With unit-norm right eigenvectors collected
in the columns of $U$ and biorthonormal left partners $\widetilde U^{\dagger}U=\mathbb{1}$,
the associated inner product is the Hermitian, positive metric
$\langle\phi,\psi\rangle=\phi^{\dagger}M\psi$ with $M=(UU^{\dagger})^{-1}$
(so that $\langle\widetilde u_m|u_n\rangle=\delta_{mn}$ holds to machine
precision while the right states remain non-orthogonal,
$|\langle u_+|u_-\rangle|\simeq0.73$). This metric is a Hermitian positive inner
product and is therefore a distinct object from the bilinear pairing
$\langle\psi_L|\psi_R\rangle$ used in the main text; the four-factor combination
below nevertheless coincides with Eq.~\eqref{eq:rate_function}, because in both
cases the dual state is the biorthogonal partner of the propagated one.
The biorthogonal Loschmidt echo per mode,
\begin{equation}
  L_k(t)=\frac{\langle\widetilde\psi(0)|\psi(t)\rangle\,\langle\widetilde\psi(t)|\psi(0)\rangle}
              {\langle\widetilde\psi(t)|\psi(t)\rangle\,\langle\widetilde\psi(0)|\psi(0)\rangle},
  \qquad
  \mathrm{LR}(t)=-\frac{1}{N}\ln\!\prod_k L_k(t),
  \label{eq:ssh_echo}
\end{equation}
with $|\psi(t)\rangle=e^{-\mathrm{i}H^f_k t}|u^{i}_{k-}\rangle$, is invariant under
independent rescalings of either state at either time and reduces to the ordinary
$|\langle\psi_0|\psi(t)\rangle|^2$ in the Hermitian limit.

\begin{figure}[t]
  \centering
  \includegraphics[width=\columnwidth]{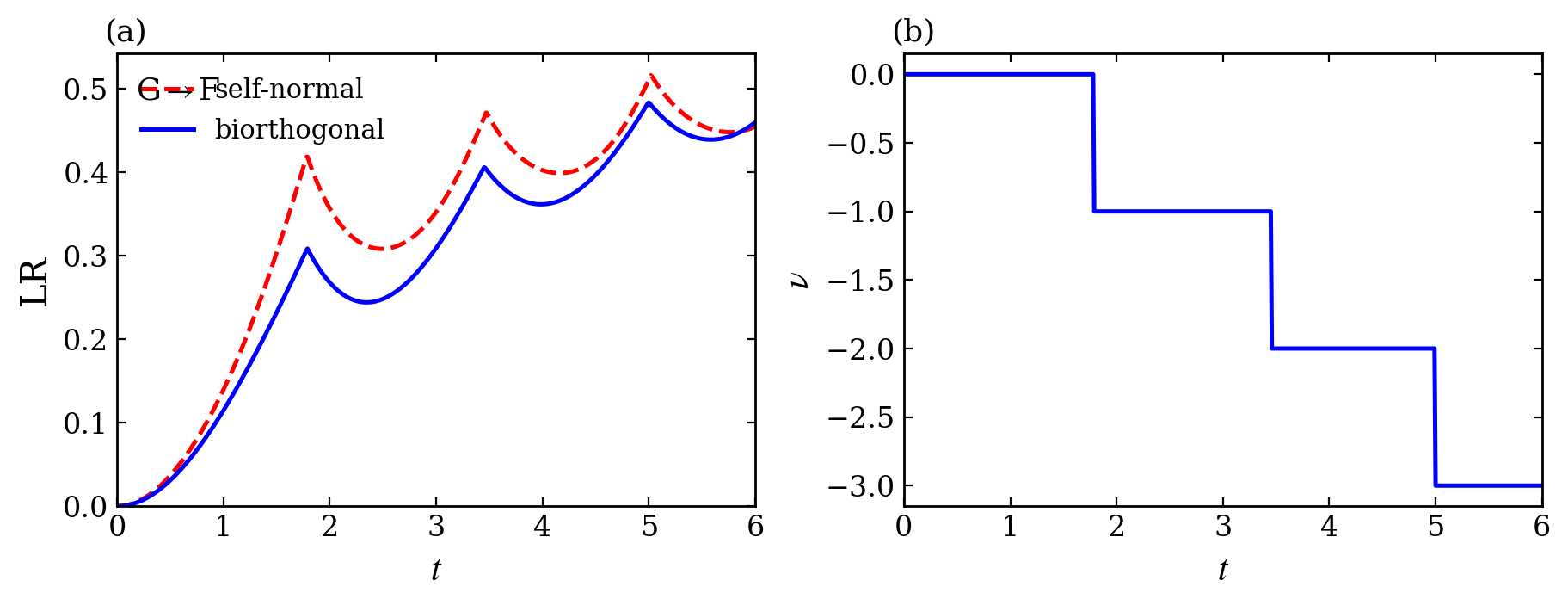}
  \caption{Biorthogonal DQPT construction validated on the non-Hermitian SSH
  model of Ref.~\onlinecite{BiorthDQPT2024}, G$\to$F quench
  [$(\eta,\gamma):(0.2,1)\to(-0.2,1)$]. (a) Loschmidt rate: the biorthogonal echo
  of Eq.~\eqref{eq:ssh_echo} (solid) lies below the self-normal rate (dashed),
  with cusps at the DQPT times. (b) Biorthogonal DTOP $\nu(t)$, which steps at the
  same critical times. Free-fermion, thermodynamic limit.}
  \label{fig:ssh_gf}
\end{figure}

\begin{figure*}[t]
  \centering
  \includegraphics[width=0.72\textwidth]{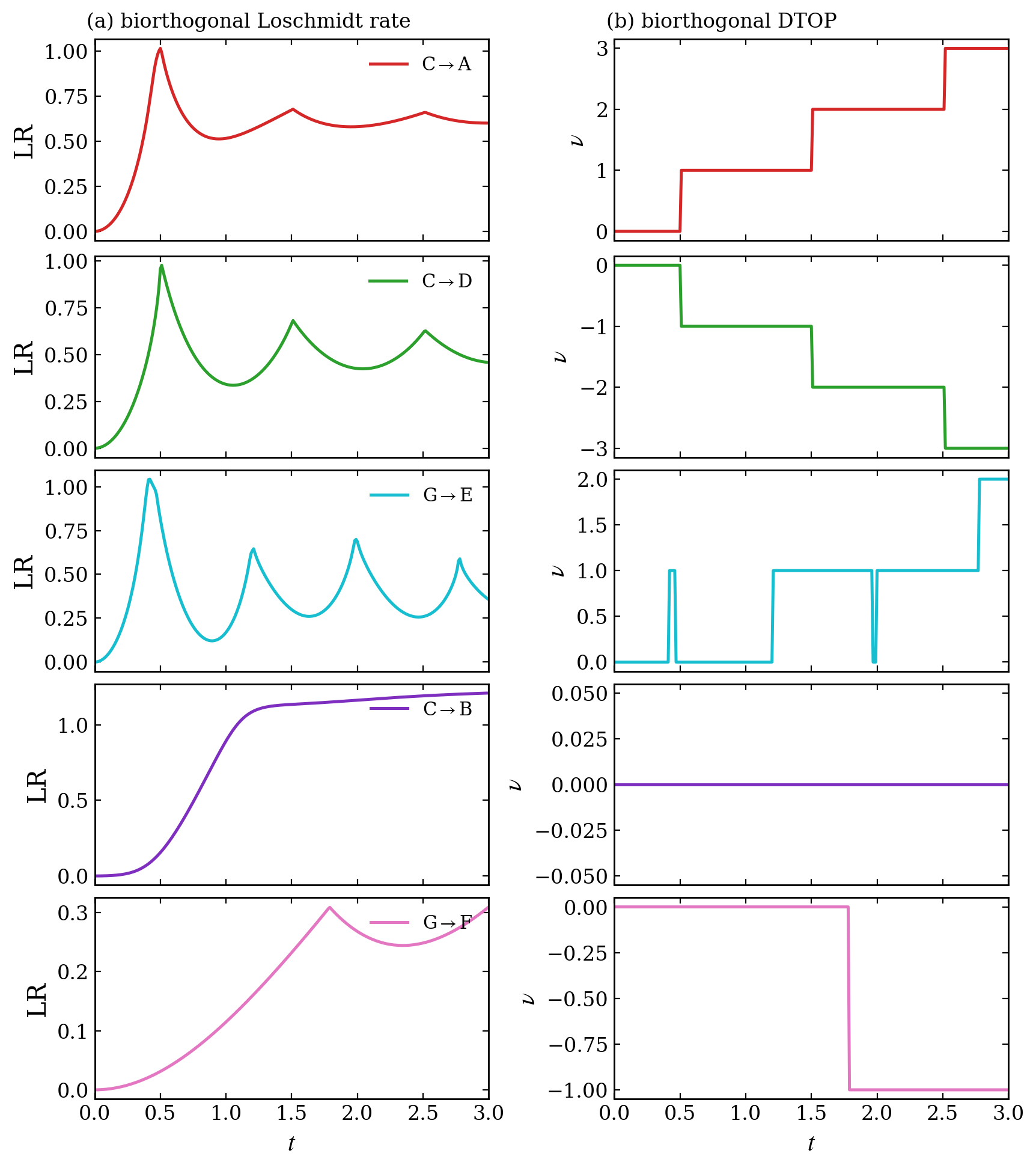}
  \caption{Five quenches of the non-Hermitian SSH model from the middle of
  the phase diagram (points as in Ref.~\onlinecite{BiorthDQPT2024}):
  C$=(0.2,5)$, A$=(-2,5)$, D$=(2,5)$, B$=(-0.2,5)$, G$=(0.2,1)$, E$=(-2,1)$,
  F$=(-0.2,1)$. (a) Biorthogonal Loschmidt rate LR$(t)$ [Eq.~\eqref{eq:ssh_echo}]
  and (b) biorthogonal DTOP $\nu(t)$. C$\to$B is a smooth crossover with no
  DQPT, and $\nu\equiv0$; all other quenches show cusps with coincident DTOP
  steps. Free-fermion, thermodynamic limit.}
  \label{fig:ssh_five}
\end{figure*}

Figure~\ref{fig:ssh_gf} reproduces the quench of
Ref.~\onlinecite{BiorthDQPT2024} between the phase-diagram points
$\mathrm{G}=(\eta,\gamma)=(0.2,1)$ and $\mathrm{F}=(-0.2,1)$: the biorthogonal rate
lies systematically below its self-normal counterpart
$-N^{-1}\ln|\langle\psi_0|\psi(t)\rangle|^2$, and both develop cusps at the DQPT
times. Figure~\ref{fig:ssh_five} shows five quenches from the ``middle'' of the
phase diagram, spanning a single dominant cusp (C$\to$A), periodic cusps
(C$\to$D, G$\to$E), a smooth crossover with \emph{no} DQPT (C$\to$B, for which the
DTOP stays zero), and a single transition (G$\to$F). The biorthogonal dynamical
topological order parameter $\nu(t)$, the winding of the geometric phase
$\phi^{G}_k=\phi_k-\phi^{\rm dyn}_k$, jumps precisely at the rate-function cusps.
Our gauge-invariant $\nu(t)$ reproduces the integer part of these changes. The
half-integer plateaus emphasized in Ref.~\onlinecite{BiorthDQPT2024} originate in
the exceptional-point branch of $\varepsilon_k$; our associated-state construction
recovers the corresponding equilibrium winding $w=\tfrac12$, but resolves the
dynamical order parameter only up to that integer part.

These curves are free-fermion and evaluated in the thermodynamic limit
($N=1500$ momenta, each mode solved exactly), and are therefore complementary to
the interacting, finite-$L$ TDVP results of the main text rather than a rerun of
them. As an exact-diagonalization check of the biorthogonal machinery itself, the
per-mode echo of Eq.~\eqref{eq:ssh_echo} evaluated through the metric $M$ agrees
with brute-force $2\times2$ propagation
$L_k(t)=|\langle\widetilde\psi(0)|\psi(t)\rangle|^2/(\langle\widetilde\psi(t)|\psi(t)\rangle\langle\widetilde\psi(0)|\psi(0)\rangle)$
to $8\times10^{-16}$, and with the closed-form single-mode expression of
Ref.~\onlinecite{BiorthDQPT2024} to $9\times10^{-13}$.

\FloatBarrier
\section{Reproducibility}
\label{app:repro}

The parameters of each numerical experiment are stated with the corresponding
model in the main text and in this Supplemental Material. Common settings, unless
stated otherwise: open boundary conditions; SVD truncation tolerance $10^{-10}$;
matrix-free Taylor tolerance $\texttt{tol}=10^{-12}$ with maximum order $M=40$;
MPO compression tolerance $10^{-8}$ with MPO bond dimension up to $256$;
normalization convention as in Sec.~\ref{app:health} (unit Schmidt vector at the
active bond after each update and truncation). Disorder, where present, is drawn
i.i.d.\ uniformly from $[-1,1]$ with the stated seed. Observables are the
center-site $\langle Z_{L/2}\rangle,\langle X_{L/2}\rangle$ for the Ising chains,
the imbalance $\mathcal{I}(t)$ for Hatano-Nelson, and the center-site
$\langle Z\rangle$ for the Lindblad example.

\section{Physical Origin of Non-Hermiticity}
\label{app:physical_origin}

The non-Hermitian terms used throughout this work are effective generators of
\emph{conditional} open-system dynamics rather than fundamental microscopic
Hamiltonians. When a Hermitian system with dissipative channels---spontaneous
emission, photoionization, inelastic loss, or engineered measurement
backaction---is conditioned on a particular measurement record, most commonly the
no-jump trajectory, the state evolves under the non-Hermitian effective generator
\begin{equation}
  H_{\rm eff} = H - \frac{i}{2}\sum_\mu L_\mu^\dagger L_\mu,
\end{equation}
which is non-Hermitian even when the microscopic couplings are not. Such
generators can also be engineered directly on digital quantum simulators through
monitored (Zeno-constrained) qudit dynamics.

This makes the model choices physically coherent. Imaginary onsite fields
$-ik\sum_i\sigma_i^z$ represent single-particle loss or decay---natural for
Rydberg arrays, trapped ions, polar molecules, and NV centers---while the
coherent long-range Ising and XY couplings are inherited from the blockade and
exchange interactions of the same platforms. Hatano-Nelson-type non-reciprocal
hopping fits the same logic but is usually realized by explicit engineering, e.g.
in photonic lattices with gain and loss, electric circuits, or dissipation-engineered
cold atoms. In all cases the microscopic world need not itself be non-Hermitian;
the experimentally relevant \emph{conditional} dynamics is.

\end{document}